\documentclass[english,aps,prd,morefloats,floatfix,showpacs]{revtex4-1}
\usepackage[T1]{fontenc}
\usepackage[latin9]{inputenc}
\usepackage{graphicx}

\makeatletter

\providecommand{\tabularnewline}{\\}

\makeatother

\usepackage{babel}

\begin{document}

\title{Effects of the FCNC couplings in production of new heavy quarks within
$Z'$ models at the LHC}

\author{V. \c{C}etinkaya}

\email{volkan.cetinkaya@dpu.edu.tr}

\selectlanguage{english}%

\affiliation{Dumlupinar University, Department of Physics, 43100 Merkez, Kutahya,
Turkey}

\author{V. Ar\i{}}

\email{vari@science.ankara.edu.tr}

\selectlanguage{english}%

\affiliation{Ankara University, Department of Physics, 06100 Tandogan, Ankara,
Turkey}

\author{O. \c{C}ak\i r}

\email{ocakir@science.ankara.edu.tr}

\selectlanguage{english}%

\affiliation{Ankara University, Department of Physics, 06100 Tandogan, Ankara,
Turkey}
\begin{abstract}
We study the flavor changing neutral current couplings of new heavy
quarks through the $Z'$ models at the LHC. We calculate the cross
sections for the signal and the corresponding standard model background
processes. Considering the present limits on the mass of new heavy
quarks and the $Z'$ boson, we performed an analysis to investigate
the parameter space (mixing and mass) through different $Z'$ models.
For an FCNC mixing parameter $x=0.1$ and the $Z'$ mass $M_{Z'}=2000$
GeV, and new heavy quark mass $m_{t'}=700$ GeV at the LHC with $\sqrt{s}=13$
TeV, we find the cross section for single production of new heavy
quarks associated with top quarks as $5.8$ fb, $3.3$ fb, $1.5$
fb and $1.2$ fb within the $Z'_{\eta}$ , $Z'_{\psi}$ , $Z'_{LP}$
and $Z'_{\chi}$ models, respectively. It is shown that the sensitivity
would benefit from the flavor tagging. 
\end{abstract}


\maketitle

\section{Introduction}

Addition of new heavy quarks would require the extension of the flavor
mixing in charged current interactions as well as the extension of
Higgs sector in the standard model (SM). A large number of new heavy
quark pairs can be produced through their colour charges at the Large
Hadron Collider (LHC). However, due to the expected smallness of the
mixing between the new heavy quarks and known quarks through charged
current interactions, the production and decay modes can be effected
by the flavor changing neutral current (FCNC) interactions. A new
symmetry beyond the SM is expected to explain the smallness of these
mixings. We may anticipate the new physics discovery by observing
large anomalous couplings in the heavy quark sector. The couplings
of the new heavy quarks can be enhanced to observable levels within
some new physics models. In numerous phenomenological studies (see
\cite{Langacker08} and references therein), a lot of extensions of
the SM foresee the extra gauge bosons, the $Z'$-boson in particular.
Flavor changing neutral currents can be induced by an extra $U(1)'$
gauge boson $Z'$. The $Z'$ boson in the models using an extra $U(1)'$
group can have tree-level or an effective $Z'q\bar{q}'$ (where $q$
and $q'$ both can be the up-type quarks or down-type quarks) couplings.
The $Z'_{\eta}$, $Z'_{\chi}$ and $Z'_{\psi}$ models corresponding
to the specific values of the mixing angle in the $E_{6}$ model with
different couplings to the fermions, and the leptophobic $Z'_{LP}$
model with the couplings to quarks but no couplings to leptons are
among the special names of the $Z'$ models \cite{Olive2014}.

The ATLAS and CMS collaborations have performed extensive searches
of new vector resonances at the LHC. We summarize briefly these searches,
that exploited data from the $pp$ run at $\sqrt{s}=7$ TeV and $\sqrt{s}=8$
TeV, as well as the corresponding constraints on $Z'$ boson masses.
The most stringent limits come from searches with leptonic final states
($Z'\to l^{+}l^{-}$): $M{}_{Z'_{\chi}}>2620$ GeV \cite{ATLAS14}
and $M{}_{Z'_{\eta}}>1870$ GeV \cite{ATLAS12}, $M_{Z'_{\psi}}>2260$
GeV \cite{CMS13} (more recently $M_{Z'_{\psi}}>2510$ GeV \cite{ATLAS14})
for the $Z'$ boson predicted by the $U(1)'$ extensions, also extending
to the mass limit of $M_{Z'_{S}}>2900$ GeV \cite{ATLAS14,CMS15}
for a gauge boson with sequential couplings. The results from ATLAS
experiment exclude a leptophobic $Z'$ decaying to $t\bar{t}$ with
a mass less than $1740$ GeV at $95\%$ C.L. \cite{Aad2013}, while
the CMS experiment excludes a top-color $Z'$ decaying to $t\bar{t}$
with a mass less than $2100$ GeV at $95\%$ C.L. \cite{Chatrchyan2013}.
These searches assume rather narrow width for the $Z'$ boson ($\Gamma_{Z'}/M_{Z'}=0.012$).
From the electroweak precision data analysis, the improved lower limits
on the $Z'$ mass are given in the range $1100-1500$ GeV, which gives
a limit on the $Z-Z'$ mixing about $10^{-3}$ \cite{Olive2014}.
The limits on the $Z'$ boson mass favors higher center of mass energy
collisions for direct observation of the signal. Using dilepton searches
with LHC data, the dark matter constraints have been analysed in Ref.
\cite{Alves2014,Alves2015} in the regime $M_{Z'}>2m_{DM}$.

A work performed in Ref. \cite{arhrib06,Cakir10} presents the effects
of FCNC interactions induced by an additional $Z'$ boson on the single
top quark and top quark pair production at the LHC ($\sqrt{s}=14$
TeV). The relevant signal cross sections have been calculated and
especially the benefit from flavor tagging to identify the signal
has been discussed. Considering an existence of sizeable couplings
to the new heavy quarks, the $Z'$ boson decay width and branchings,
as well as the production rates, can be quite different from the expectations
of usual search scenarios.

In the models of interest new heavy quarks can have some mixing with
the SM quarks. For example, in composite Higgs model \cite{Contino03}
the lightest new heavy quark couples predominantly to the heavier
SM quarks (top and bottom quarks). In the models of vector-like quarks
(VLQ) \cite{Aguilar-Saavedra13} they are expected to couple preferentially
to third-generation quarks and they can have flavour-changing neutral
current couplings, in addition to the charged-current decays characteristic
of chiral quarks. Within the $E_{6}$ model the isosinglet quarks
\cite{Sultansoy08} are predicted and they can decay to the quarks
of the SM. The new heavy quarks can be produced dominantly in pairs
through strong interactions for masses around 1 TeV in the $pp$ collisions
of the LHC with a center of mass energy of $13$ TeV. The single production
of new heavy quarks would only be dominant over pair production for
the large quark masses \cite{OCakir2008}, it is model dependent,
and it could be suppressed if the mixing with SM quarks is small.

There are searches for pair production and single production of new
heavy quarks at the LHC. The ATLAS and CMS collaborations focused
on decay modes of new heavy quarks into a massive vector boson and
a third generation quark assuming a $100\%$ branching ratio, based
on $L_{int}\approx20$ fb$^{-1}$ of $pp$ collision data at $\sqrt{s}=8$
TeV, and set lower mass limit for up type new heavy quark as $m_{t'}>700$
GeV \cite{CMS2014A} and $m_{t'}>735$ GeV \cite{ATLAS2014AZ}.

In this work, we investigate the single production of new heavy quarks
via FCNC interactions through $Z'$ boson exchange at the LHC. This
paper aims at studying the signal and background in detail within
the same MC framework, and the relevant interaction vertices are implemented
into the MC software. Analyzing the signal observability (via contour
plots) for different mass values of the $Z'$ boson and new heavy
quarks as well as the mixing parameter through FCNC interactions are
another feature of the work. In section II, we calculate the decay
widths and branching ratios of $Z^{\prime}$ boson for the mass range
$1500-3000$ GeV in the framework of different $Z'$ models. An analysis
of the parameter space of mass and coupling strength is given for
the single production of new heavy quarks at the LHC in section III.
We analyzed the signal observability for the $Z'q\bar{q}'$ FCNC interactions.
We consider both $t'\bar{t}$ and $\bar{t}'t$ single new heavy quark
productions for the purpose of enriching the signal statistics even
at the small couplings. The analysis for the signal significance is
given in section IV and the work ends up with the conclusions as given
in section V.

\section{FCNC interactions}

In the gauge eigenstate basis, following the formalism given in Ref.
\cite{arhrib06,langacker00,cheung07}, the additional neutral current
Lagrangian associated with the $U(1)^{'}$ gauge symmetry can be written
as 
\begin{equation}
\mathcal{L}'=-g'\sum_{f,f'}\bar{f}\gamma^{\mu}\left[\epsilon_{L}(ff')P_{L}+\epsilon_{R}(ff')P_{R}\right]f'Z_{\mu}^{'}
\end{equation}
where $\epsilon_{L,R}(ff')$ are the chiral couplings of $Z'$ boson
with fermions $f$ and $f'$. The $g'$ is the gauge coupling of the
$U(1)^{'}$, and $P_{R,L}=(1\pm\gamma^{5})/2.$ Here, we assume that
there is no mixing between the $Z$ and $Z'$ bosons as favored by
the precision data. Flavor changing neutral currents (FCNCs) arise
if the chiral couplings are nondiagonal matrices. In case the $Z'$
couplings are diagonal but nonuniversal, flavor changing couplings
are emerged by fermion mixing. In the interaction basis the FCNC for
the up-type quarks are given by

\begin{equation}
\mathcal{J}_{FCNC}^{u}=\left(\overline{u},\overline{c},\overline{t},\overline{t}'\right)\gamma_{\mu}(\epsilon_{L}^{u}P_{L}+\epsilon_{R}^{u}P_{R})\left(\begin{array}{c}
u\\
c\\
t\\
t'
\end{array}\right),
\end{equation}
where the chiral couplings are given by

\begin{equation}
\epsilon_{L}^{u}=C_{L}^{u}\left(\begin{array}{cccc}
1 & 0 & 0 & 0\\
0 & 1 & 0 & 0\\
0 & 0 & 1 & 0\\
0 & 0 & 0 & x
\end{array}\right)\quad\mbox{and}\quad\epsilon_{R}^{u}=C_{R}^{u}\left(\begin{array}{cccc}
1 & 0 & 0 & 0\\
0 & 1 & 0 & 0\\
0 & 0 & 1 & 0\\
0 & 0 & 0 & 1
\end{array}\right).
\end{equation}

In general, the effects of these FCNCs may occur both in the up-type
sector and down-type sector after diagonalizing their mass matrices.
For the right-handed up-sector and down-sector one assumes that the
neutral current couplings to $Z'$ are family universal and flavor
diagonal in the interaction basis. In this case, unitary rotations
($V_{L,R}^{f}$) can keep the right handed couplings flavor diagonal,
and left handed sector becomes nondiagonal. The chiral couplings of
$Z'$ in the fermion mass eigenstate basis are given by

\begin{equation}
B_{L}^{ff'}\equiv V_{L}^{f}\epsilon_{L}^{u}(ff')V_{L}^{f^{\dagger}}\quad\mbox{and}\quad B_{R}^{ff'}\equiv V_{R}^{f}\epsilon_{R}^{u}(ff')V_{R}^{f^{\dagger}}.
\end{equation}
Here the matrix can be written as $V'=V_{L}^{u}V_{L}^{d\dagger}$
with the assumption that the down-sector has no mixing. The flavor
mixing in the left-handed quark fields is simply related to $V'$,
assuming the up sector diagonalization and unitarity of the CKM matrix
one can find the couplings $B_{L}^{u}\equiv V_{uL}^{\dagger}\epsilon_{L}^{u}V_{uL}=V'\epsilon_{L}^{u}V'^{\dagger}$
with the parametrization $|V_{iQ}|=|A_{iQ}|\lambda^{4-i}$ for the
matrix, where the generation index $i$ runs from 1 to 3.

The FCNC effects from the $Z'$ mediation have been studied for the
down-type sector and implications in flavor physics through $B$-meson
decays \cite{barger04,Barger:2004hn,Chen:2006vs,barger09,barger09-1,Alok2010}
and $B$-meson mixing \cite{cheung07,Leroux:2001fx,Barger:2004qc,He:2006bk,Chiang:2006we,Baek:2006bv}.
These effects have also been studied for up-type quark sector in top
quark production \cite{arhrib06,CorderoCid:2005kp,Yue:2003wd,Lee:2000km,Yue:2006qd,delAguila:2009gz}.
The parameters for different $Z'$ models are given in Table \ref{tab:tab1}.
In numerical calculations, we take the coupling $g'\simeq0.40$ for
the models.

In our model, the chiral couplings can be written as

\begin{equation}
B_{L}^{u}\approx C_{L}^{u}\left(\begin{array}{cccc}
1+\left(x-1\right)|A_{14}|^{2}\lambda^{6} & \left(x-1\right)A_{14}A_{24}^{*}\lambda^{5} & \left(x-1\right)A_{14}A_{34}^{*}\lambda^{4} & \left(x-1\right)A_{14}A_{44}^{*}\lambda^{3}\\
\left(x-1\right)A_{24}A_{14}^{*}\lambda^{5} & 1+\left(x-1\right)|A_{24}|^{2}\lambda^{4} & \left(x-1\right)A_{24}A_{34}^{*}\lambda^{3} & \left(x-1\right)A_{24}A_{44}^{*}\lambda^{2}\\
\left(x-1\right)A_{34}A_{14}^{*}\lambda^{4} & \left(x-1\right)A_{34}A_{24}^{*}\lambda^{3} & 1+\left(x-1\right)|A_{34}|^{2}\lambda^{2} & \left(x-1\right)A_{34}A_{44}^{*}\lambda^{1}\\
\left(x-1\right)A_{44}A_{14}^{*}\lambda^{3} & \left(x-1\right)A_{44}A_{24}^{*}\lambda^{2} & \left(x-1\right)A_{44}A_{34}^{*}\lambda^{1} & 1+\left(x-1\right)|A_{44}|^{2}
\end{array}\right)
\end{equation}

The values of the matrix elements $|A_{14}|=3.2$, $|A_{24}|=2.0$
and $|A_{34}|=3.0$ are used as given in Ref. \cite{Bobrowski} by
taking into account $\lambda=0.22$. For a comparison, we also calculate
the cross sections using the scenario of equal parameters $|A_{i4}|=2.0$
(where $i$ runs from $1$ to $3$). 

\begin{table}
\caption{The chiral couplings of $Z'$ boson with quarks and leptons predicted
by different models.\label{tab:tab1}}

\begin{tabular}{|c|c|c|c|c|}
\hline 
 & $Z'_{\chi}$  & $Z'_{\psi}$  & $Z'_{\eta}$  & $Z'_{LP}$\tabularnewline
\hline 
\hline 
$C_{L}^{u}$  & $-1/2\sqrt{10}$  & $1/\sqrt{24}$  & $-1/\sqrt{15}$  & $1/9$\tabularnewline
\hline 
$C_{R}^{u}$  & $1/2\sqrt{10}$  & $-1/\sqrt{24}$  & $1/\sqrt{15}$  & $-8/9$\tabularnewline
\hline 
$C_{L}^{d}$  & $-1/2\sqrt{10}$  & $1/\sqrt{24}$  & $-1/\sqrt{15}$  & $1/9$\tabularnewline
\hline 
$C_{R}^{d}$  & $-3/2\sqrt{10}$  & $-1/\sqrt{24}$  & $-1/2\sqrt{15}$  & $1/9$\tabularnewline
\hline 
$C_{L}^{e}$  & $3/2\sqrt{10}$  & $1/\sqrt{24}$  & $1/2\sqrt{15}$  & $0$\tabularnewline
\hline 
$C_{R}^{e}$  & $1/2\sqrt{10}$  & $-1/\sqrt{24}$  & $1/\sqrt{15}$  & $0$\tabularnewline
\hline 
$C_{L}^{\nu}$  & $3/2\sqrt{10}$  & $1/\sqrt{24}$  & $1/2\sqrt{15}$  & $0$\tabularnewline
\hline 
\end{tabular}
\end{table}

For the FCNC constraints from $D^{0}-\bar{D}^{0}$ mixing with parameter
$x_{D}\approx6\times10^{-3}$, we follow the calculations performed
in Ref. \cite{arhrib06}, and find that the contribution from $Z'$
boson (through FCNC effects) can be obtained as $x_{D}^{Z'}\simeq2\times10^{4}(C_{L}^{uc})^{2}$,
where $C_{L}^{uc}=C_{L}^{u}(x-1)|A_{14}||A_{24}|\lambda^{5}$. With
the given parametrizations above, this is translated into the result
that as long as the combination $|C_{L}^{u}(x-1)|$ is less than about
$0.3$, the experimental bounds can be well satisfied in different
$Z'$ models that we have studied in this work.

\section{Single production of new heavy quarks}

\begin{figure}
\includegraphics[scale=0.8]{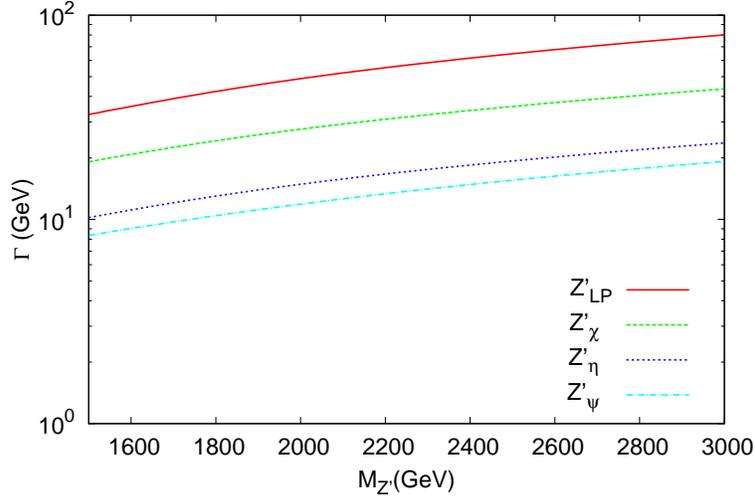}

\caption{The decay widths of $Z'$ boson depending on its mass for different
$Z'$ models with the FCNC parameter $x=0.1$. Here, we use the mass
value of new heavy quarks as 700 GeV, and the mass value of new heavy
charged lepton and neutrino as 200 GeV and 100 GeV, respectively.
\label{fig:1}}
\end{figure}

\begin{figure}
\includegraphics{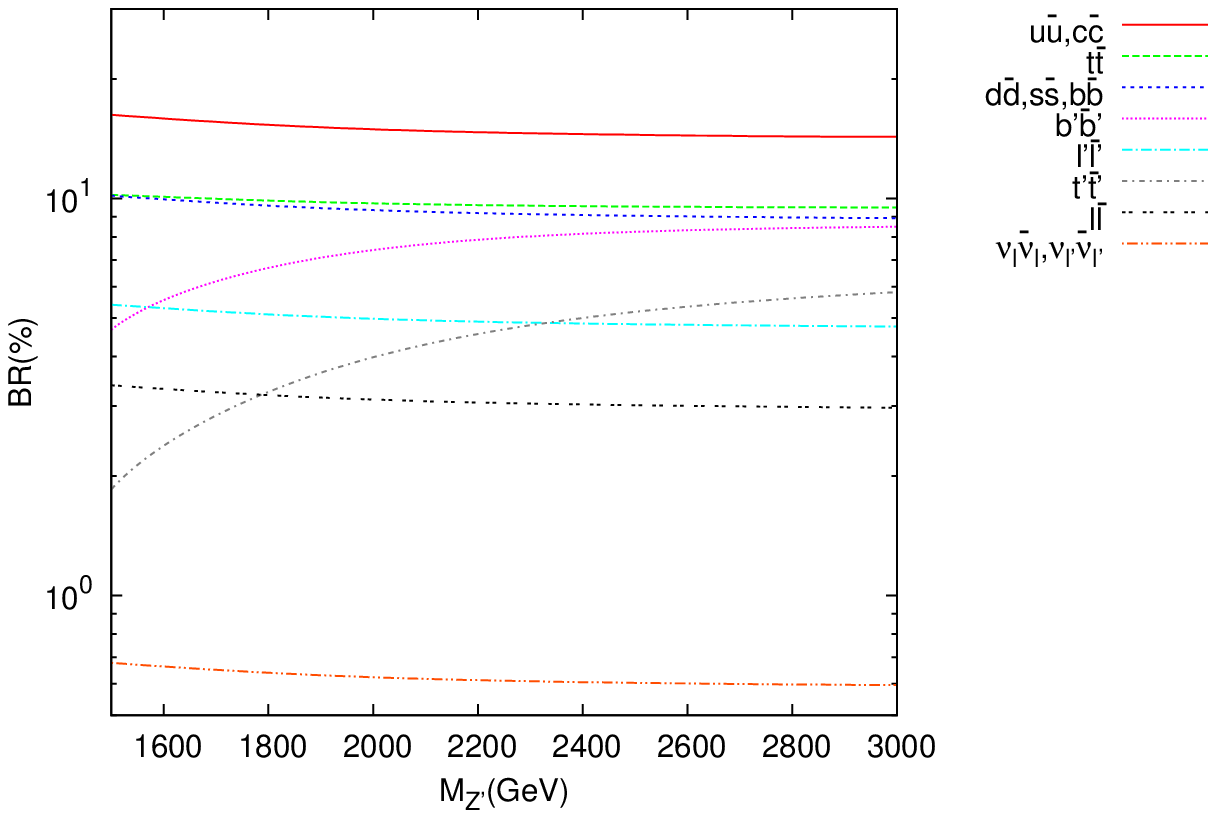}

\caption{Branching ratios (\%) depending on the mass of $Z'$ boson for diagonal
couplings to quarks and leptons within the $Z'_{\eta}$ model. The
new heavy quark masses are taken to be 700 GeV, and new heavy charged
lepton mass is 200 GeV and new heavy neutrino mass is 100 GeV.\label{fig:fig2}}
\end{figure}

\begin{figure}
\includegraphics{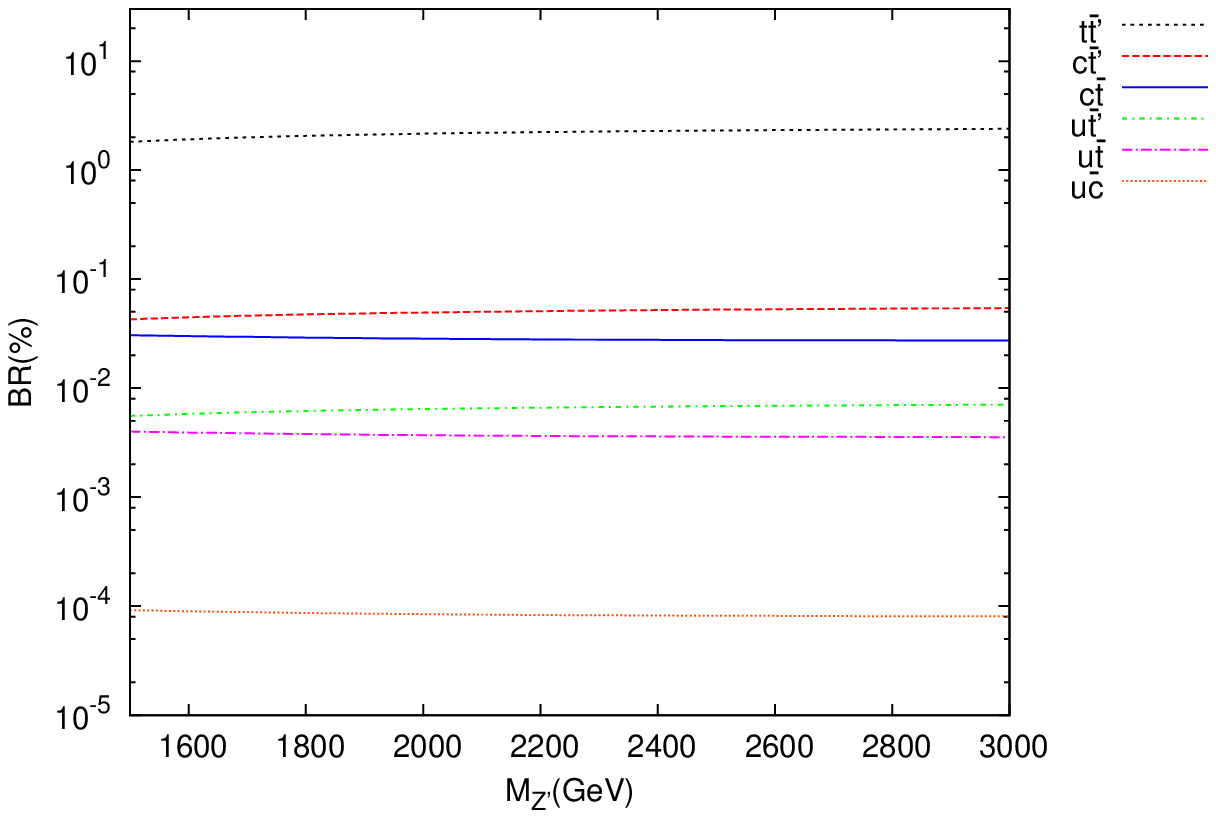}

\caption{Branching ratios (\%) depending on the mass of the $Z'$ boson for
FCNC couplings to different flavors of up sector quarks within the
$Z'_{\eta}$ model. Here the mass of $t'$ quark is 700 GeV and the
FCNC parameter $x=0.1$.\label{fig:fig3}}
\end{figure}

\begin{figure}
\includegraphics{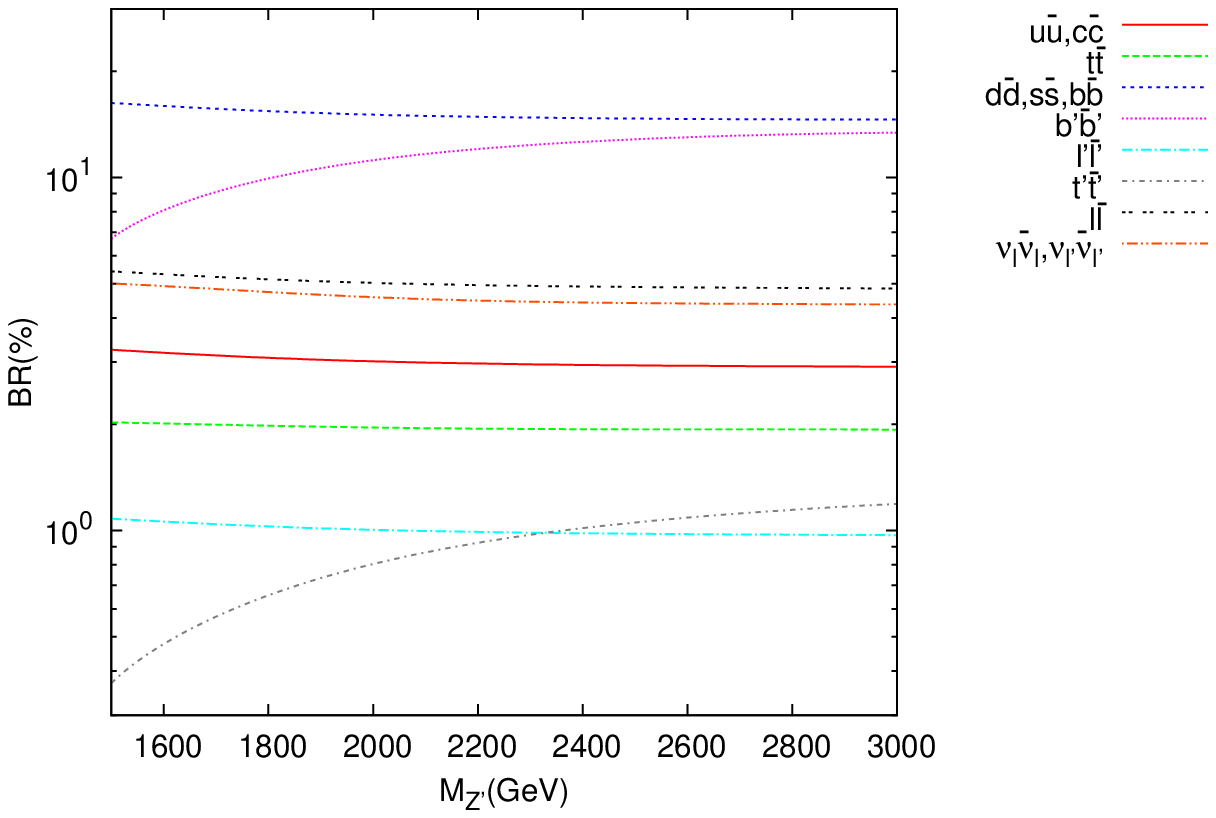}

\caption{Branching ratios (\%) depending on the mass of $Z'$ boson for diagonal
couplings to quarks and leptons within the $Z'_{\chi}$ model. The
new heavy quark masses are taken to be 700 GeV, and new heavy charged
lepton mass is 200 GeV and new heavy neutrino mass is 100 GeV. \label{fig:fig4}}
\end{figure}

\begin{figure}
\includegraphics{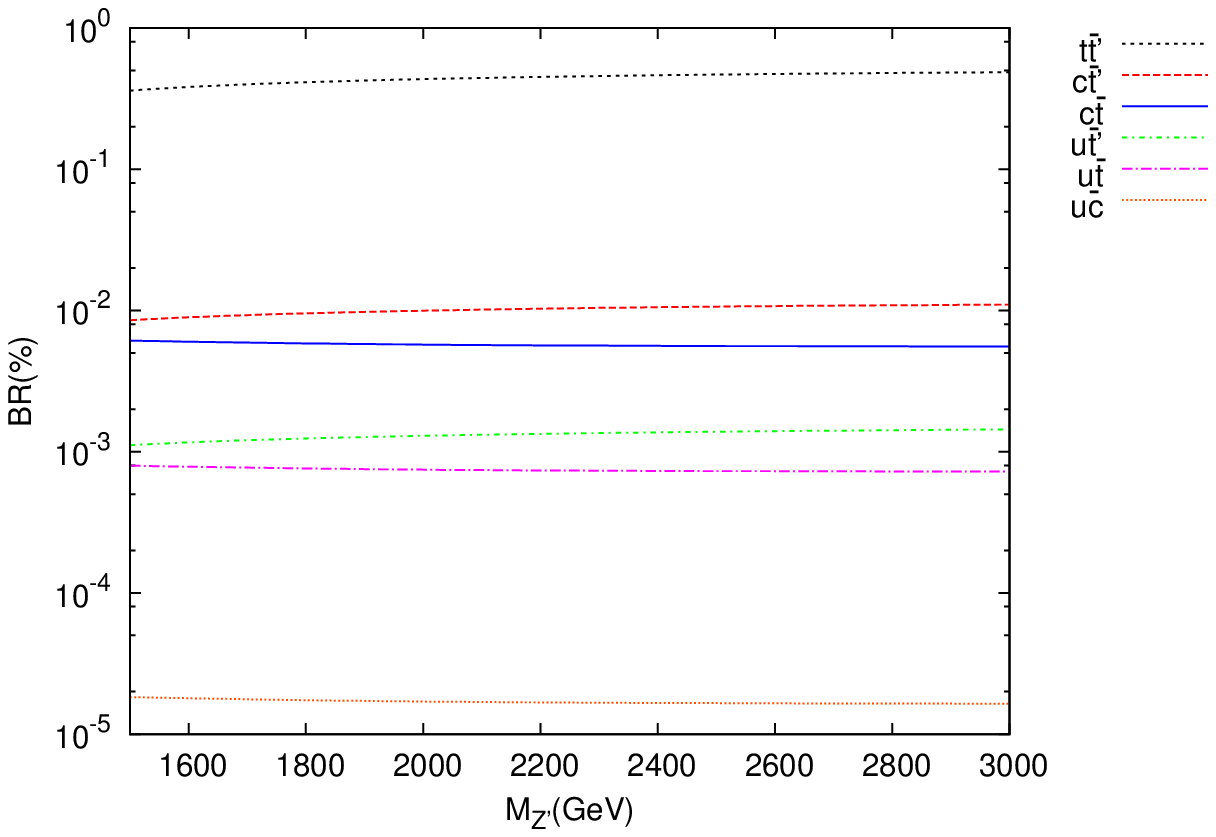}

\caption{Branching ratios (\%) depending on the mass of the $Z'$ boson for
FCNC couplings to different flavors of up sector quarks within the
$Z'_{\chi}$ model. Here the mass of $t'$ quark is 700 GeV and the
FCNC parameter $x=0.1$.\label{fig:fig5}}
\end{figure}

\begin{figure}
\includegraphics{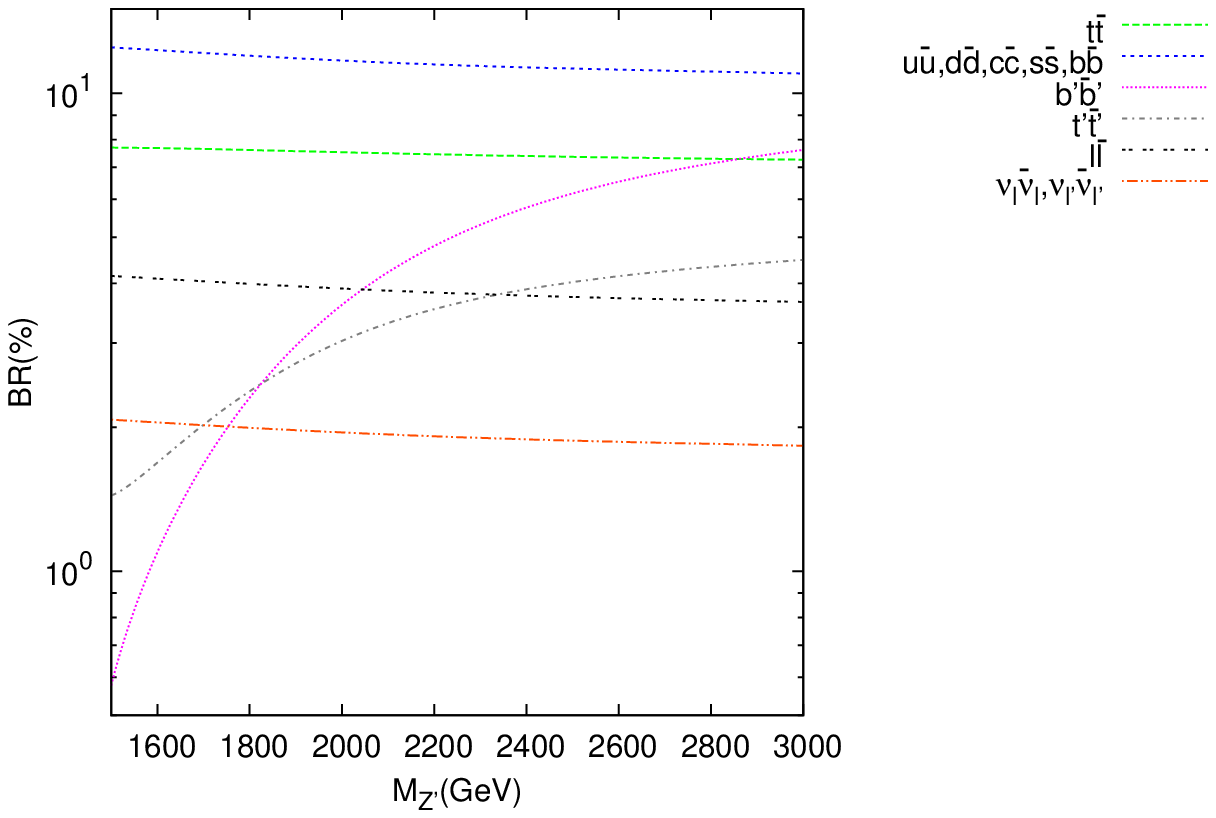}

\caption{Branching ratios (\%) depending on the mass of $Z'$ boson for diagonal
couplings to quarks and leptons within the $Z'_{\psi}$ model. The
new heavy quark masses are taken to be 700 GeV, and new heavy charged
lepton mass is 200 GeV and new heavy neutrino mass is 100 GeV. \label{fig:fig6}}
\end{figure}

\begin{figure}
\includegraphics{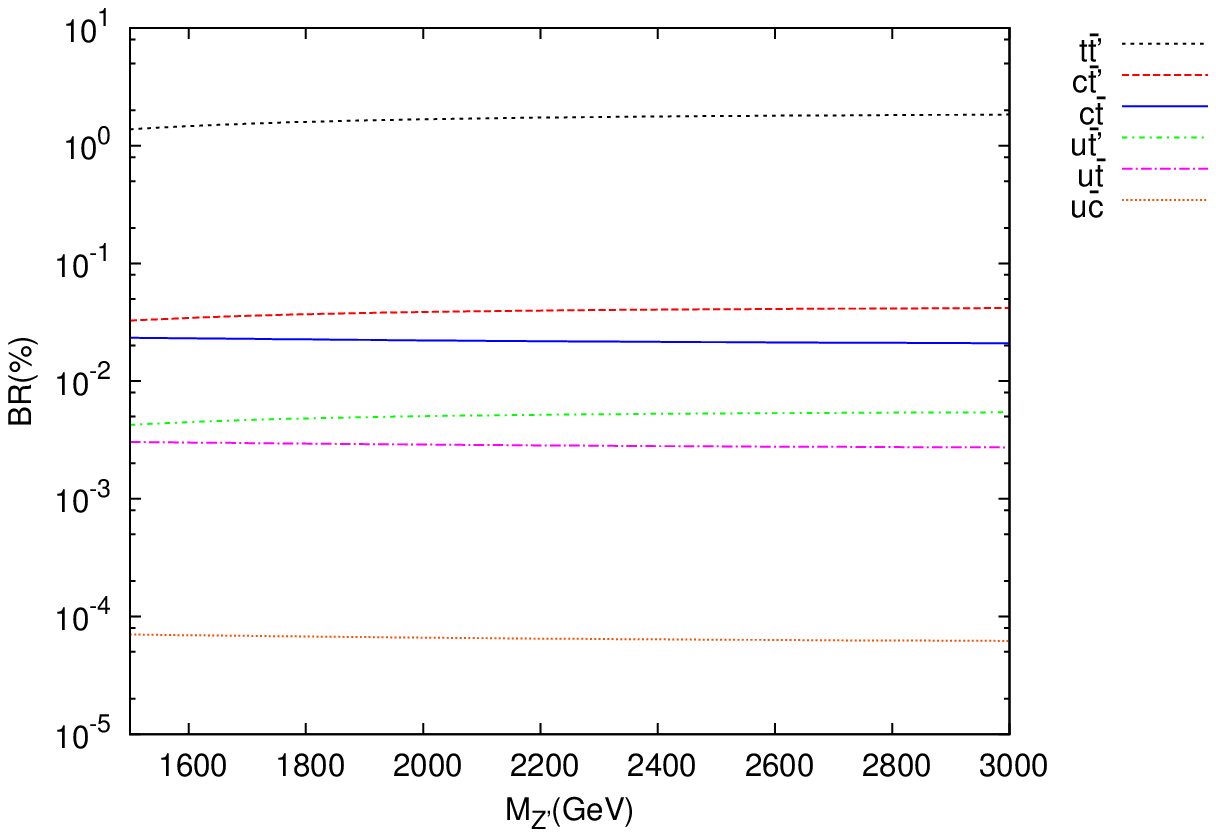}

\caption{Branching ratios (\%) depending on the mass of the $Z'$ boson for
FCNC couplings to different flavors of up sector quarks within the
$Z'_{\psi}$ model. Here the mass of $t'$ quark is 700 GeV and the
FCNC parameter $x=0.1$. \label{fig:fig7}}
\end{figure}

\begin{figure}
\includegraphics{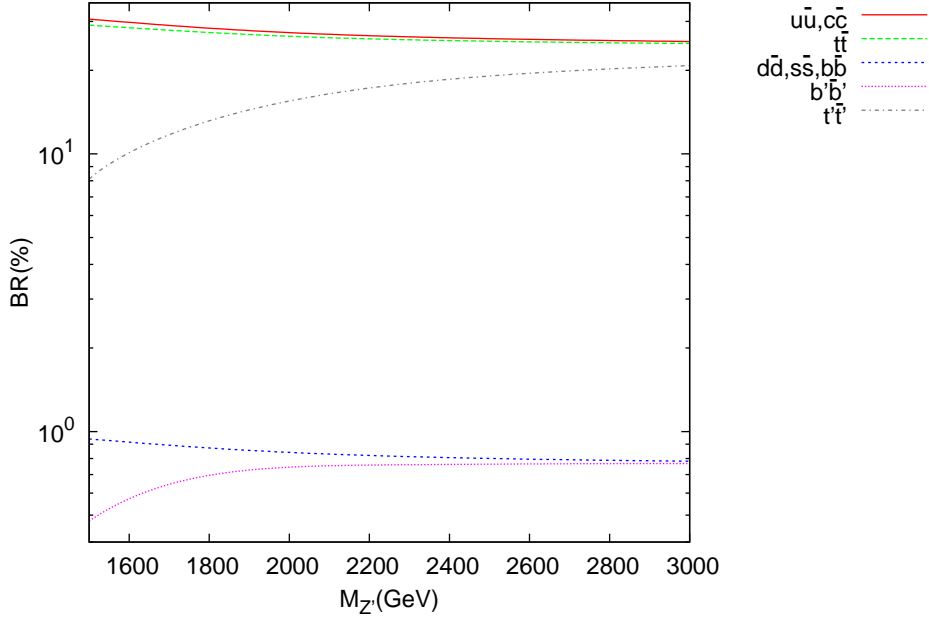}

\caption{Branching ratios (\%) depending on the mass of $Z'$ boson for diagonal
couplings to quarks within the $Z'_{LP}$ model. The new heavy quark
masses are taken to be 700 GeV. \label{fig:fig8} }
\end{figure}

\begin{figure}
\includegraphics{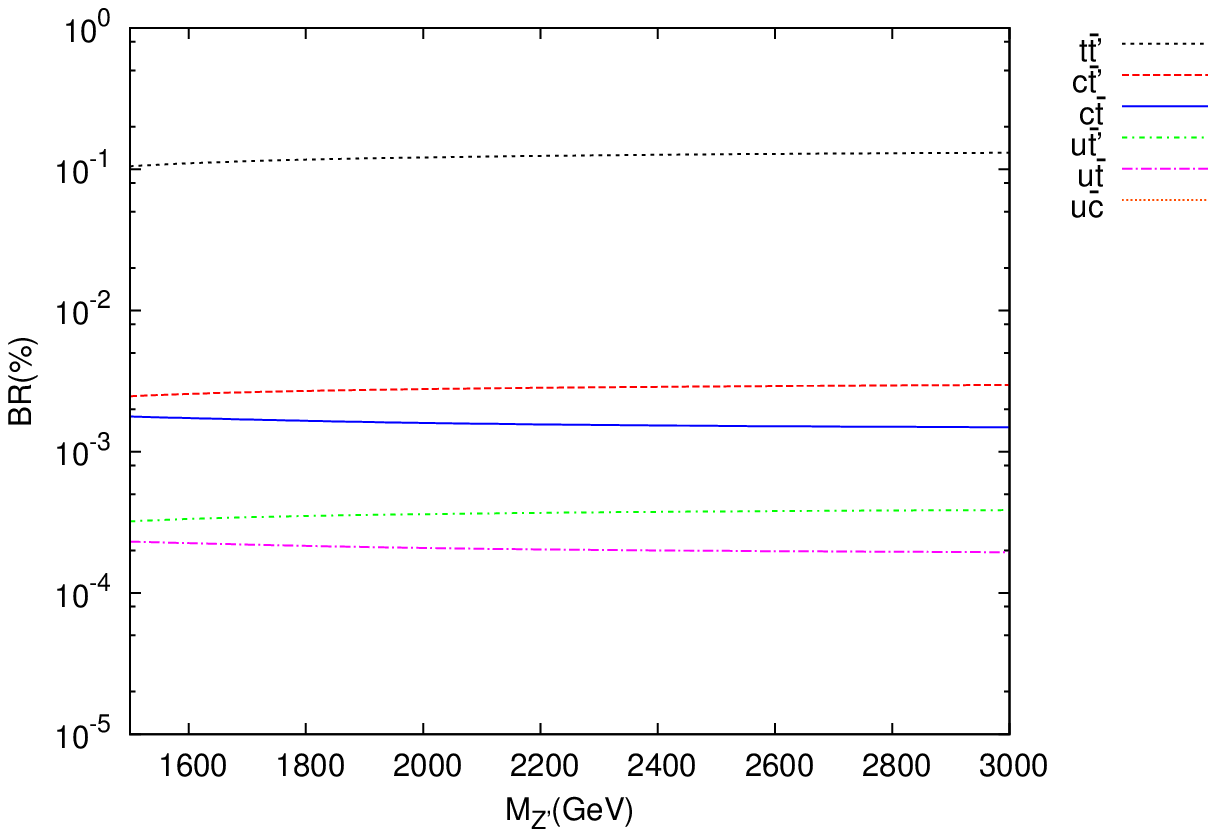}

\caption{Branching ratios (\%) depending on the mass of the $Z'$ boson for
FCNC couplings to different flavors of up sector quarks within the
$Z'_{LP}$ model. Here the mass of $t'$ quark is 700 GeV and the
FCNC parameter $x=0.1$.\label{fig:fig9}}
\end{figure}

For numerical calculations we have implemented the $Z^{\prime}qq'$
interaction vertices into the CalcHEP program package \cite{Pukhov2013}.
The decay widths of $Z'$ boson for different mass values within different
$Z'$ models are given in Table \ref{tab:tab2}. For the parameter
$x=1$, both the left-handed and right-handed couplings become universal,
and family diagonal. In this case we cannot see the FCNC effects on
the decay widths and cross sections. For the FCNC effects on the decay
width, we take the parameter $x=0.1$ as shown in Fig. \ref{fig:1}.
All these scenarios of $Z'$ models predict a narrow decay width ranging
from $0.6\%$ to $3\%$ for $\Gamma_{Z'}/M_{Z'}$ depending on the
mass of $Z'$ boson foreseen by different models, for the considered
set of parameters. The effect of the FCNC reduces the decay width
in the relevant mass range. The decay widths are compared with the
similar results from Ref. \cite{arhrib06,Cakir10} for $x=0.1$ to
prove the implementation. Unless otherwise stated thoughout this work,
we use the FCNC mixing parameter $x=0.1$ and the mass value of $t'$
quark $m_{t'}=700$ GeV, and the mass value of new heavy charged lepton
$m_{l'}=200$ GeV and new heavy neutrino $m_{\nu'}=100$ GeV. The
branching ratios of $Z'$ boson decays depending on its mass predicted
by different $Z'$ models are given in Fig. \ref{fig:fig2} - Fig.
\ref{fig:fig9}, specifically they are given in Fig. \ref{fig:fig2},
Fig. \ref{fig:fig4} and Fig. \ref{fig:fig6} for the diagonal couplings
to quarks and leptons, while they are given in Fig. \ref{fig:fig3},
Fig. \ref{fig:fig5} and Fig. \ref{fig:fig7} for the FCNC couplings
to different flavors of up sector quarks within the $Z'_{\eta}$,
$Z'_{\chi}$ and $Z'_{\psi}$ models, respectively. In Fig. \ref{fig:fig8}
and Fig. \ref{fig:fig9}, the branchings for a leptophobic $Z'_{LP}$
boson decays to pair of quarks with diagonal couplings and FCNC couplings
are presented depending on its mass.

\begin{table}
\caption{The total decay widths of $Z'$ boson for different mass values in
various models with the FCNC parameter $x=0.1$. We use the value
of masses of new heavy quarks $m_{Q'}=700$ GeV, and the masses of
new heavy leptons $m_{l'}=200$ GeV and $m_{\nu'}=100$ GeV.\label{tab:tab2}}

\begin{tabular}{|c|c|c|c|c|}
\hline 
$M_{Z'}(\mbox{GeV})$  & $\Gamma(Z'_{\chi})(\mbox{GeV})$  & $\Gamma(Z'_{\psi})(\mbox{GeV})$  & $\Gamma(Z'_{\eta})(\mbox{GeV})$  & $\Gamma(Z'_{LP})(\mbox{GeV})$\tabularnewline
\hline 
\hline 
1400  & 17.25  & 7.73  & 9.21  & 28.85\tabularnewline
\hline 
1600  & 21.75  & 9.20  & 11.61  & 37.44\tabularnewline
\hline 
1800  & 25.11  & 10.66  & 13.46  & 43.99\tabularnewline
\hline 
2000  & 28.30  & 12.11  & 15.22  & 50.25\tabularnewline
\hline 
2200  & 31.41  & 13.55  & 16.95  & 56.36\tabularnewline
\hline 
2400  & 34.48  & 14.98  & 18.65  & 62.36\tabularnewline
\hline 
2600  & 37.53  & 16.40  & 20.33  & 68.30\tabularnewline
\hline 
2800  & 40.55  & 17.82  & 22.01  & 74.18\tabularnewline
\hline 
3000  & 43.56  & 19.23  & 23.67  & 80.02\tabularnewline
\hline 
\end{tabular}
\end{table}

\begin{figure}
\includegraphics{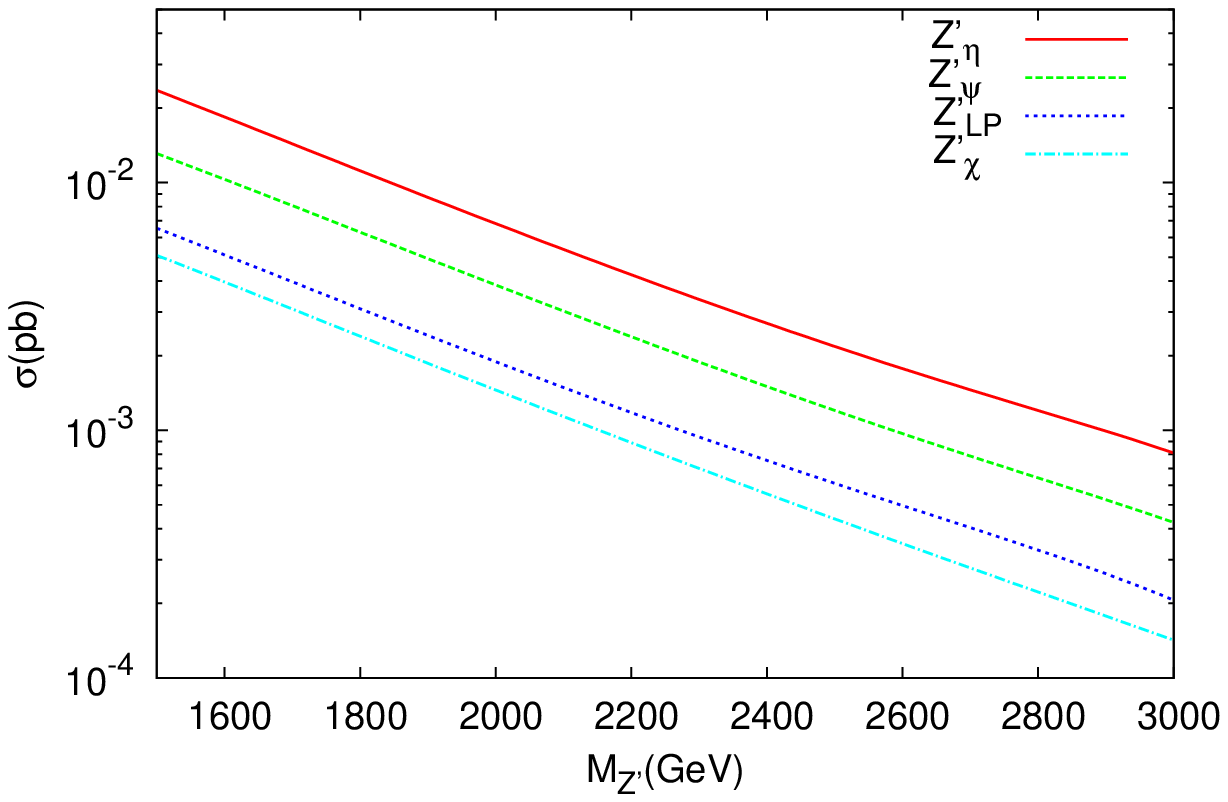}

\caption{The cross sections for $pp\to(t'\bar{t}+\bar{t}'t)+X$ versus the
$Z'$ boson mass at the LHC with $\sqrt{s}=$13 TeV. The lines are
for different $Z'$ models as explained in the text, where the parameters
$|A_{14}|=3.2$, $|A_{24}|=2.0$ and $|A_{34}|=3.0$ are used. The
mass value of $t'$ quark is used as $m_{t'}=700$ GeV and the FCNC
parameter is used as $x=0.1$. \label{fig:10}}
\end{figure}

\begin{figure}
\includegraphics{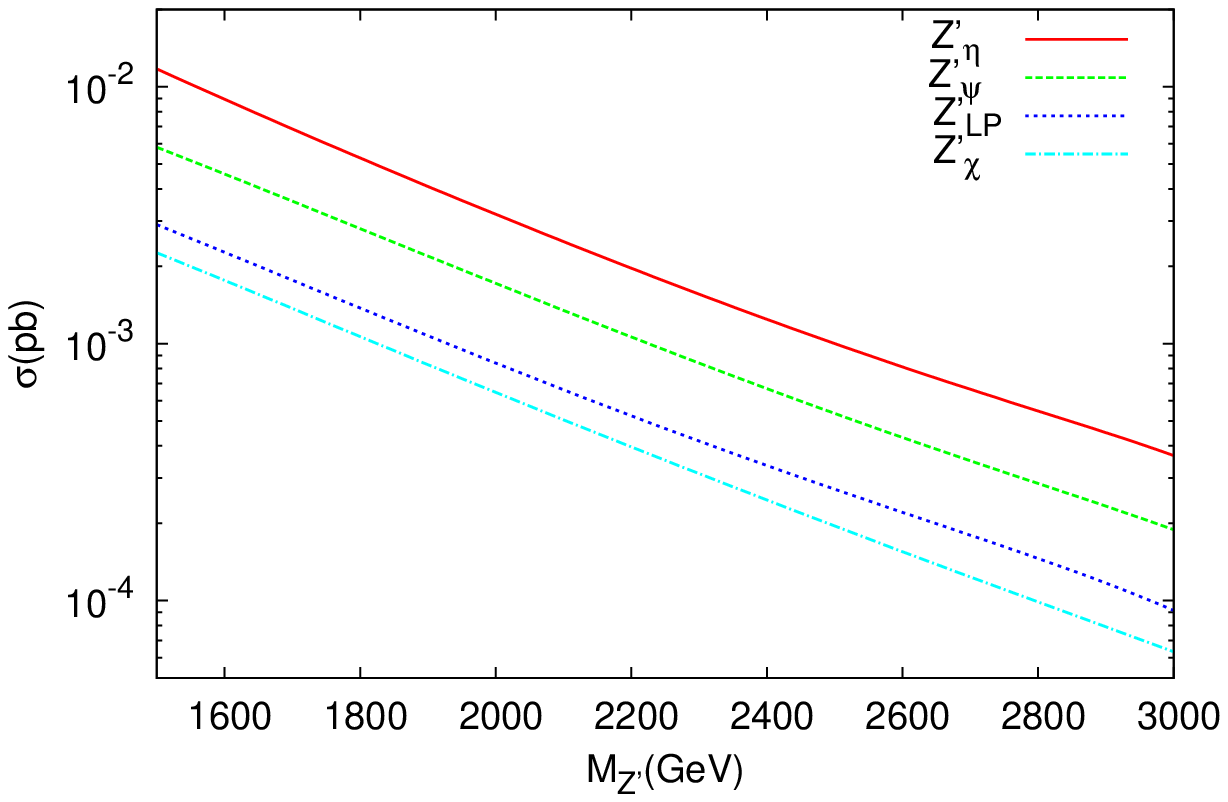}

\caption{The cross sections for $pp\to(t'\bar{t}+\bar{t}'t)+X$ versus the
$Z'$ boson mass at the LHC with $\sqrt{s}=$13 TeV. The lines are
for the four $Z'$ models as explained in the text and the values
of parameters $|A_{i4}|=2$ (where $i$ runs form 1 to 3) are used.
The mass value of $t'$ quark is used as $m_{t'}=700$ GeV and the
FCNC parameter is used as $x=0.1$. \label{fig:11}}
\end{figure}

The cross sections for the process $pp\to(t'\bar{t}+\bar{t}'t)+X$
depending on the $Z'$ boson mass at the LHC ($\sqrt{s}=$13 TeV)
are given in Fig. \ref{fig:10} and Fig. \ref{fig:11} by using parton
distribution function library CTEQ6L \cite{Pumplin:2002vw}. Here,
the $Z'$ boson contributes through the $s$- and $t$-channel diagrams,
and the cross sections of associated production of single top quarks
and single new heavy quarks ($t'\bar{t}$ and $\bar{t}'t$) in the
final state are summed. For this process the cross section at $\sqrt{s}=13$
TeV is about 8 times larger than the case at $\sqrt{s}=8$ TeV.

\begin{figure}
\includegraphics[scale=0.7]{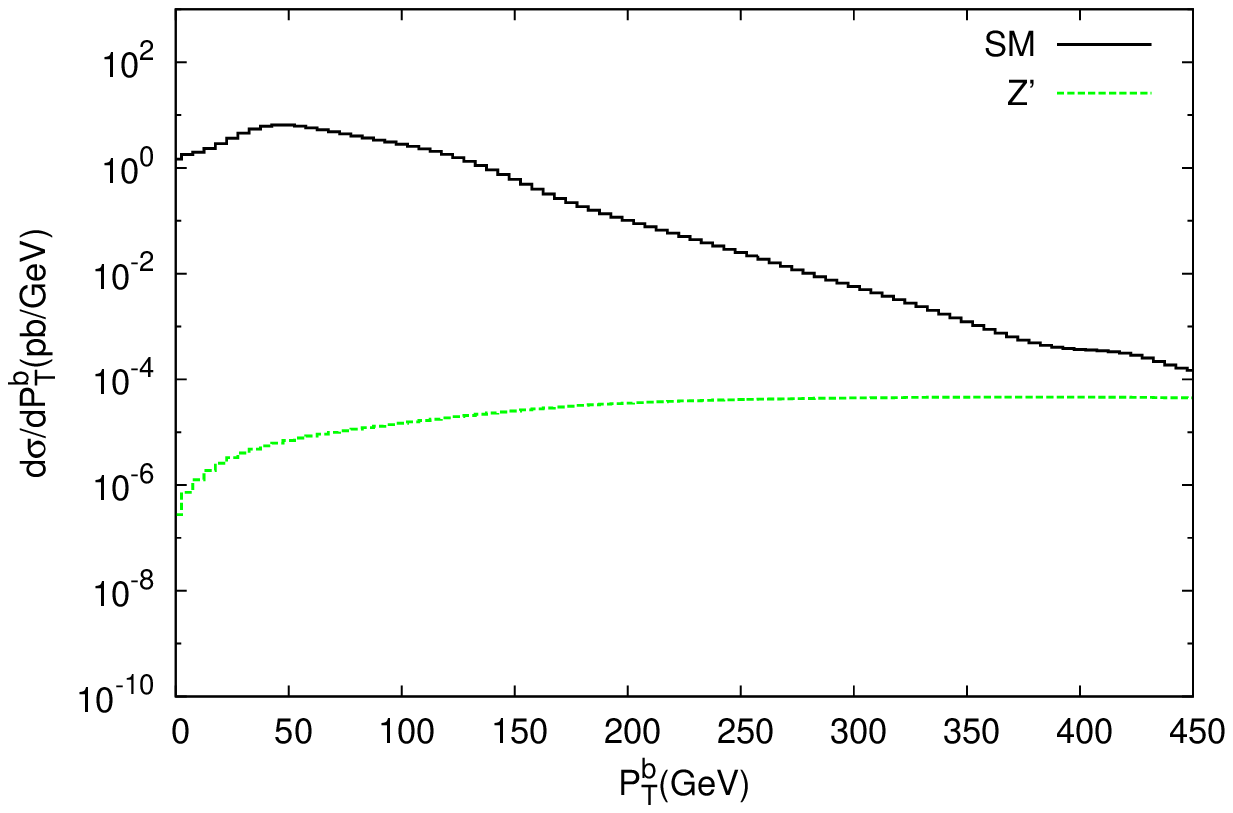}

\caption{The transverse momentum ($p_{T}$) distribution of the bottom quarks
($b$ and $\bar{b}$) for the signal and background processes $pp\rightarrow\left(W^{+}b\bar{t}+W^{-}\bar{b}t\right)+X$
at the LHC with $\sqrt{s}=13$ TeV for the parameters explained in
the text. These results are obtained for the $Z'_{\eta}$ model. \label{fig:12}}
\end{figure}

\begin{figure}
\includegraphics[scale=0.7]{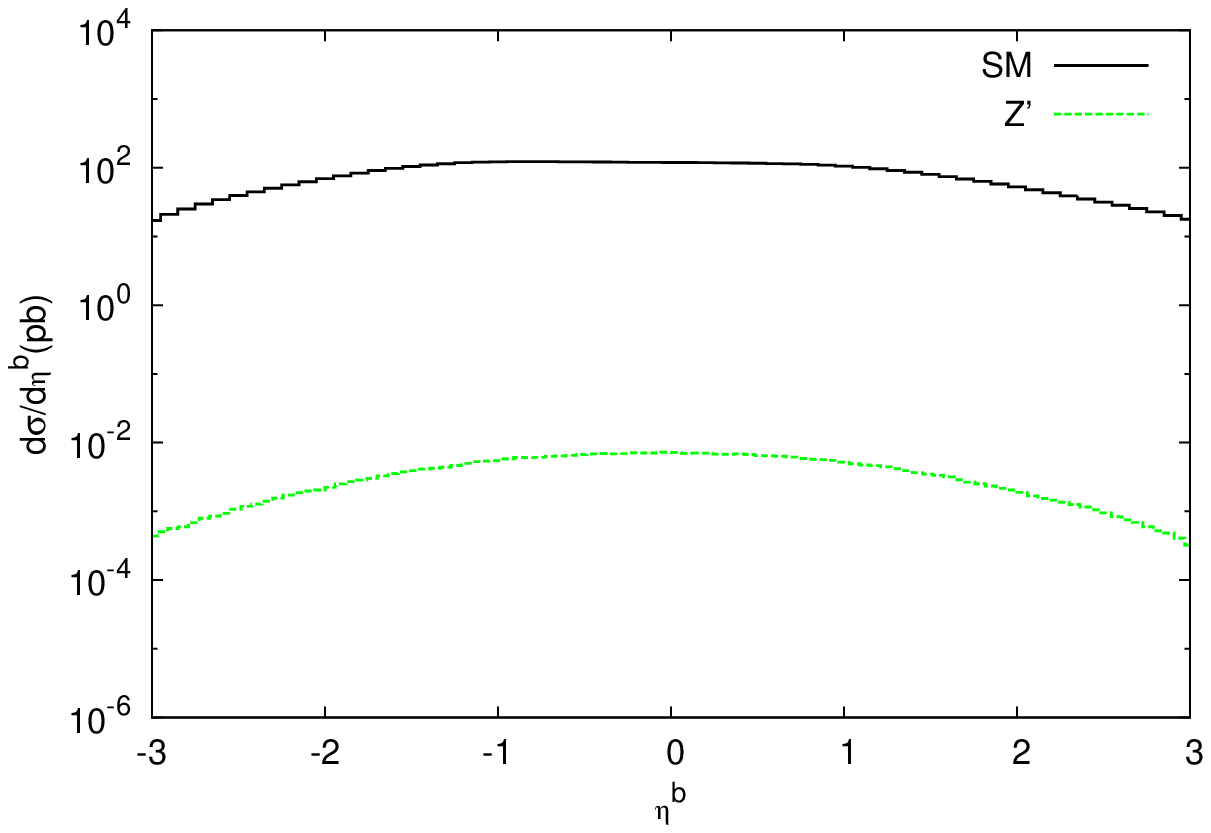}

\caption{The pseudo-rapidity distribution of the bottom quarks ($b$ and $\bar{b}$)
for the signal ($Z'$) and background processes (SM) $pp\rightarrow\left(W^{+}b\bar{t}+W^{-}\bar{b}t\right)+X$
at the LHC with $\sqrt{s}=13$ TeV for the parameters explained in
the text. The results are obtained for the $Z'_{\eta}$ model. \label{fig:13} }
\end{figure}

Fig. \ref{fig:12} shows the $p_{T}$ distributions of the $b$-quark
in the signal process with $M_{Z^{'}}=1500$ GeV for the parameter
$x=0.1$ at the $pp$ center of mass energy of 13 TeV. A high $p_{T}$
cut reduces the background significantly without affecting much the
signal cross section in the interested $Z'$ mass range. The rapidity
distribution of the bottom quarks ($b$ and $\bar{b}$) from the signal
are shown in Fig. \ref{fig:13} at the collision energy of 13 TeV.
In order to enhance the statistics we sum up the $b$ and $\bar{b}$
distributions. There is a peak in the $b$-quark rapidity distribution
$\eta^{b}\simeq0$ with the tails extending to $|\eta^{b}|\simeq2.5$.
For the analysis, the suitable cuts are $p_{T}^{b,\bar{b}}>100$ GeV,
$|\eta^{b,\bar{b}}|\leq2.0$ and $m_{Wb}>400$ GeV. The cut $m_{Wb}$
is also useful well above the top quark mass. We also apply invariant
mass cut (for $Wbt$ system) \textit{\emph{$M_{Z'}-4\Gamma_{Z'}<m_{Wbt}<M_{Z'}+4\Gamma$}}
to make analysis with the signal and background.

The signal cross sections \textit{\emph{($\sigma_{S}$) in the invariant
mass interval of $M_{Z'}-4\Gamma_{Z'}<m_{Wbt}<M_{Z'}+4\Gamma$,}}
for the process $pp\rightarrow\left(W^{+}b\bar{t}+W^{-}\bar{b}t\right)+X$
are given in Table \ref{tab:tab3}, Table \ref{tab:tab4}, Table \ref{tab:tab5}
and Table \ref{tab:tab6} for different values of FCNC parameter $x$
(ranging from 0.01 to 0.5) within the $Z'_{\eta}$ ($Z'_{\chi}$)
model, where new heavy quark masses are taken 600 GeV, 700 GeV and
800 GeV. For the $Z'_{\psi}$ ($Z'_{LP}$) model and the FCNC parameter
for $0.01$, $0.05$, $0.1$ and $0.5$, the cross sections ($\sigma_{S}$)
are presented in Table \ref{tab:tab7}, Table \ref{tab:tab8}, Table
\ref{tab:tab9} and Table \ref{tab:tab10}, respectively. The cross
section for the corresponding background are given in Table \ref{tab:tab11}
for the chosen invariant mass interval.

\begin{table}
\caption{The total cross section for the signal ($\sigma_{S}$) depending on
the $Z'$ boson mass, these values are calculated for the process
$pp\rightarrow\left(W^{+}b\bar{t}+W^{-}\bar{b}t\right)+X$ at the
LHC with $\sqrt{s}=13$ TeV. The results are obtained for the $Z'_{\eta}$
model (the $Z'_{\chi}$ model) with the FCNC parameter $x=0.01$.\label{tab:tab3}}

\begin{tabular}{|c|c|c|c|}
\hline 
\multicolumn{1}{|c|}{} & \multicolumn{3}{c|}{$\sigma_{S}\left(\mbox{pb}\right)$}\tabularnewline
\hline 
$M_{Z'}(\mbox{GeV})$  & $m_{t'}=600\mbox{ GeV}$  & $m_{t'}=700\mbox{ GeV}$  & $m_{t'}=800\mbox{ GeV}$\tabularnewline
\hline 
$1500$  & $2.44\times10^{-2}$ ($5.51\times10^{-3}$)  & $2.23\times10^{-2}$ ($4.98\times10^{-3}$)  & $1.98\times10^{-2}$ ($4.41\times10^{-3}$)\tabularnewline
\hline 
$2000$  & $7.31\times10^{-3}$ ($1.60\times10^{-3}$)  & $6.95\times10^{-3}$ ($1.51\times10^{-3}$)  & $6.49\times10^{-3}$ ($1.41\times10^{-3}$)\tabularnewline
\hline 
$2500$  & $2.40\times10^{-3}$ ($5.05\times10^{-4}$)  & $2.33\times10^{-3}$ ($4.90\times10^{-4}$)  & $2.24\times10^{-3}$ ($4.70\times10^{-4}$)\tabularnewline
\hline 
$3000$  & $1.01\times10^{-3}$ ($1.74\times10^{-4}$)  & $9.93\times10^{-4}$ ($1.71\times10^{-4}$)  & $9.71\times10^{-4}$ ($1.67\times10^{-4}$)\tabularnewline
\hline 
\end{tabular}
\end{table}

\begin{table}
\caption{The same as Table \ref{tab:tab3}, but for $x=0.05$. \label{tab:tab4}}

\begin{tabular}{|c|c|c|c|}
\hline 
\multicolumn{1}{|c|}{} & \multicolumn{3}{c|}{$\sigma_{S}\left(\mbox{pb}\right)$}\tabularnewline
\hline 
$M_{Z'}(\mbox{GeV})$  & $m_{t'}=600\mbox{ GeV}$  & $m_{t'}=700\mbox{ GeV}$  & $m_{t'}=800\mbox{ GeV}$\tabularnewline
\hline 
$1500$  & $2.26\times10^{-2}$ ($5.06\times10^{-3}$)  & $2.06\times10^{-2}$ ($4.59\times10^{-3}$)  & $1.82\times10^{-2}$ ($4.07\times10^{-3}$)\tabularnewline
\hline 
$2000$  & $6.75\times10^{-3}$ ($1.47\times10^{-3}$)  & $6.44\times10^{-3}$ ($1.40\times10^{-3}$)  & $6.00\times10^{-3}$ ($1.29\times10^{-3}$)\tabularnewline
\hline 
$2500$  & $2.22\times10^{-3}$ ($4.65\times10^{-4}$)  & $2.16\times10^{-3}$ ($4.51\times10^{-4}$)  & $2.07\times10^{-3}$ ($4.32\times10^{-4}$)\tabularnewline
\hline 
$3000$  & $9.28\times10^{-4}$ ($1.60\times10^{-4}$)  & $9.17\times10^{-4}$ ($1.58\times10^{-4}$)  & $8.96\times10^{-4}$ ($1.54\times10^{-4}$)\tabularnewline
\hline 
\end{tabular}
\end{table}

\begin{table}
\caption{The same as Table \ref{tab:tab3}, but for $x=0.1$.\label{tab:tab5}}

\begin{tabular}{|c|c|c|c|}
\hline 
\multicolumn{1}{|c|}{} & \multicolumn{3}{c|}{$\sigma_{S}\left(\mbox{pb}\right)$}\tabularnewline
\hline 
$M_{Z'}(\mbox{GeV})$  & $m_{t'}=600\mbox{ GeV}$  & $m_{t'}=700\mbox{ GeV}$  & $m_{t'}=800\mbox{ GeV}$\tabularnewline
\hline 
$1500$  & $2.04\times10^{-2}$ ($4.55\times10^{-3}$)  & $1.86\times10^{-2}$ ($4.12\times10^{-3}$)  & $1.64\times10^{-2}$ ($3.66\times10^{-3}$)\tabularnewline
\hline 
$2000$  & $6.10\times10^{-3}$ ($1.32\times10^{-3}$)  & $5.81\times10^{-3}$ ($1.25\times10^{-3}$)  & $5.42\times10^{-3}$ ($1.16\times10^{-3}$)\tabularnewline
\hline 
$2500$  & $2.00\times10^{-3}$ ($4.19\times10^{-4}$)  & $1.95\times10^{-3}$ ($4.05\times10^{-4}$)  & $1.87\times10^{-3}$ ($3.89\times10^{-4}$)\tabularnewline
\hline 
$3000$  & $8.34\times10^{-4}$ ($1.44\times10^{-4}$)  & $8.23\times10^{-4}$ ($1.42\times10^{-4}$)  & $8.04\times10^{-4}$ ($1.38\times10^{-4}$)\tabularnewline
\hline 
\end{tabular}
\end{table}

\begin{table}
\caption{The same as Table \ref{tab:tab3}, but for $x=0.5$.\label{tab:tab6}}

\begin{tabular}{|c|c|c|c|}
\hline 
\multicolumn{1}{|c|}{} & \multicolumn{3}{c|}{$\sigma_{S}\left(\mbox{pb}\right)$}\tabularnewline
\hline 
$M_{Z'}(\mbox{GeV})$  & $m_{t'}=600\mbox{ GeV}$  & $m_{t'}=700\mbox{ GeV}$  & $m_{t'}=800\mbox{ GeV}$\tabularnewline
\hline 
$1500$  & $6.45\times10^{-3}$ ($1.41\times10^{-3}$)  & $5.85\times10^{-3}$ ($1.28\times10^{-3}$)  & $5.09\times10^{-3}$ ($1.13\times10^{-3}$)\tabularnewline
\hline 
$2000$  & $1.91\times10^{-3}$ ($4.09\times10^{-4}$)  & $1.82\times10^{-3}$ ($3.89\times10^{-4}$)  & $1.71\times10^{-3}$ ($3.61\times10^{-4}$)\tabularnewline
\hline 
$2500$  & $6.22\times10^{-4}$ ($1.30\times10^{-4}$)  & $6.09\times10^{-4}$ ($1.26\times10^{-4}$)  & $5.87\times10^{-4}$ ($1.20\times10^{-4}$)\tabularnewline
\hline 
$3000$  & $2.55\times10^{-4}$ ($4.43\times10^{-5}$)  & $2.51\times10^{-4}$ ($4.36\times10^{-5}$)  & $2.47\times10^{-4}$ ($4.26\times10^{-5}$)\tabularnewline
\hline 
\end{tabular}
\end{table}

\begin{table}
\caption{The total cross section values for the signal ($\sigma_{S}$) are
calculated for the process $pp\rightarrow\left(W^{+}b\bar{t}+W^{-}\bar{b}t\right)+X$
at the LHC with $\sqrt{s}=13$ TeV. The results are obtained for the
$Z'_{\psi}$ model (the $Z'_{LP}$ model) and the FCNC parameter $x=0.01$.\label{tab:tab7}}

\begin{tabular}{|c|c|c|c|}
\hline 
\multicolumn{1}{|c|}{} & \multicolumn{3}{c|}{$\sigma_{S}\left(\mbox{pb}\right)$}\tabularnewline
\hline 
$M_{Z'}(\mbox{GeV})$  & $m_{t'}=600\mbox{ GeV}$  & $m_{t'}=700\mbox{ GeV}$  & $m_{t'}=800\mbox{ GeV}$\tabularnewline
\hline 
$1500$  & $1.42\times10^{-2}$ ($6.22\times10^{-3}$)  & $1.28\times10^{-2}$ ($5.77\times10^{-3}$)  & $1.08\times10^{-2}$ ($5.20\times10^{-3}$)\tabularnewline
\hline 
$2000$  & $4.13\times10^{-3}$ ($1.84\times10^{-3}$)  & $3.97\times10^{-3}$ ($1.77\times10^{-3}$)  & $3.72\times10^{-3}$ ($1.67\times10^{-3}$)\tabularnewline
\hline 
$2500$  & $1.33\times10^{-3}$ ($6.13\times10^{-4}$)  & $1.30\times10^{-3}$ ($6.01\times10^{-4}$)  & $1.26\times10^{-3}$ ($5.82\times10^{-4}$)\tabularnewline
\hline 
$3000$  & $5.24\times10^{-4}$ ($3.00\times10^{-4}$)  & $5.20\times10^{-4}$ ($3.00\times10^{-4}$)  & $5.13\times10^{-4}$ ($2.95\times10^{-4}$)\tabularnewline
\hline 
\end{tabular}
\end{table}

\begin{table}
\caption{The same as Table \ref{tab:tab7}, but for $x=0.05$. \label{tab:tab8}}

\begin{tabular}{|c|c|c|c|}
\hline 
\multicolumn{1}{|c|}{} & \multicolumn{3}{c|}{$\sigma_{S}\left(\mbox{pb}\right)$}\tabularnewline
\hline 
$M_{Z'}(\mbox{GeV})$  & $m_{t'}=600\mbox{ GeV}$  & $m_{t'}=700\mbox{ GeV}$  & $m_{t'}=800\mbox{ GeV}$\tabularnewline
\hline 
$1500$  & $1.31\times10^{-2}$ ($5.72\times10^{-3}$)  & $1.18\times10^{-2}$ ($5.33\times10^{-3}$)  & $9.90\times10^{-3}$ ($4.79\times10^{-3}$)\tabularnewline
\hline 
$2000$  & $3.82\times10^{-3}$ ($1.70\times10^{-3}$)  & $3.67\times10^{-3}$ ($1.64\times10^{-3}$)  & $3.43\times10^{-3}$ ($1.55\times10^{-3}$)\tabularnewline
\hline 
$2500$  & $1.23\times10^{-3}$ ($5.64\times10^{-4}$)  & $1.20\times10^{-3}$ ($5.54\times10^{-4}$)  & $1.17\times10^{-3}$ ($5.35\times10^{-4}$)\tabularnewline
\hline 
$3000$  & $4.84\times10^{-4}$ ($1.77\times10^{-4}$)  & $5.21\times10^{-4}$ ($2.76\times10^{-4}$)  & $4.72\times10^{-4}$ ($2.72\times10^{-4}$)\tabularnewline
\hline 
\end{tabular}
\end{table}

\begin{table}
\caption{The same as Table \ref{tab:tab7}, but for $x=0.1$.\label{tab:tab9}}

\begin{tabular}{|c|c|c|c|}
\hline 
\multicolumn{1}{|c|}{} & \multicolumn{3}{c|}{$\sigma_{S}\left(\mbox{pb}\right)$}\tabularnewline
\hline 
$M_{Z'}(\mbox{GeV})$  & $m_{t'}=600\mbox{ GeV}$  & $m_{t'}=700\mbox{ GeV}$  & $m_{t'}=800\mbox{ GeV}$\tabularnewline
\hline 
$1500$  & $1.18\times10^{-2}$ ($5.17\times10^{-3}$)  & $1.06\times10^{-2}$ ($4.78\times10^{-3}$)  & $8.93\times10^{-3}$ ($4.30\times10^{-3}$)\tabularnewline
\hline 
$2000$  & $3.43\times10^{-3}$ ($1.53\times10^{-3}$)  & $3.30\times10^{-3}$ ($1.47\times10^{-3}$)  & $3.10\times10^{-3}$ ($1.39\times10^{-3}$)\tabularnewline
\hline 
$2500$  & $1.11\times10^{-3}$ ($5.06\times10^{-4}$)  & $1.09\times10^{-3}$ ($4.97\times10^{-4}$)  & $1.05\times10^{-3}$ ($4.82\times10^{-4}$)\tabularnewline
\hline 
$3000$  & $4.33\times10^{-4}$ ($2.49\times10^{-4}$)  & $4.31\times10^{-4}$ ($2.48\times10^{-4}$)  & $4.24\times10^{-4}$ ($2.44\times10^{-4}$)\tabularnewline
\hline 
\end{tabular}
\end{table}

\begin{table}
\caption{The same as Table \ref{tab:tab7}, but for $x=0.5$.\label{tab:tab10}}

\begin{tabular}{|c|c|c|c|}
\hline 
\multicolumn{1}{|c|}{} & \multicolumn{3}{c|}{$\sigma_{S}\left(\mbox{pb}\right)$}\tabularnewline
\hline 
$M_{Z'}(\mbox{GeV})$  & $m_{t'}=600\mbox{ GeV}$  & $m_{t'}=700\mbox{ GeV}$  & $m_{t'}=800\mbox{ GeV}$\tabularnewline
\hline 
$1500$  & $3.72\times10^{-3}$ ($1.60\times10^{-3}$)  & $3.34\times10^{-3}$ ($1.49\times10^{-3}$)  & $2.78\times10^{-3}$ ($1.33\times10^{-3}$)\tabularnewline
\hline 
$2000$  & $1.08\times10^{-3}$ ($4.74\times10^{-4}$)  & $1.04\times10^{-3}$ ($4.57\times10^{-4}$)  & $9.72\times10^{-4}$ ($4.31\times10^{-4}$)\tabularnewline
\hline 
$2500$  & $3.44\times10^{-4}$ ($1.57\times10^{-4}$)  & $3.39\times10^{-4}$ ($1.54\times10^{-4}$)  & $3.29\times10^{-4}$ ($1.49\times10^{-4}$)\tabularnewline
\hline 
$3000$  & $1.33\times10^{-4}$ ($7.69\times10^{-5}$)  & $1.32\times10^{-4}$ ($7.66\times10^{-5}$)  & $1.30\times10^{-4}$ ($7.57\times10^{-5}$)\tabularnewline
\hline 
\end{tabular}
\end{table}

\begin{table}
\caption{The cross section for the background ($\sigma_{B}$) depending on
the values of invariant mass interval, the values are calculated for
the process $pp\rightarrow\left(W^{+}b\bar{t}+W^{-}\bar{b}t\right)X$
at the LHC with $\sqrt{s}=13$ TeV. The results are given in the invariant
mass between $M_{Z'}-4\Gamma_{Z'}<m_{Wbt}<M_{Z'}+4\Gamma_{Z'}$ for
different $Z'$ decay widths corresponding to the parameters explained
in the text. \label{tab:tab11}}

\begin{tabular}{|c|c|c|c|c|}
\hline 
 & \multicolumn{4}{c|}{$\sigma_{B}\left(\mbox{pb}\right)$}\tabularnewline
\hline 
$m_{Wbt}(\mbox{GeV})$  & (for $\Gamma_{Z'_{\eta}}$ width)  & (for $\Gamma_{Z'_{\chi}}$width)  & (for $\Gamma_{Z'_{\psi}}$width)  & (for $\Gamma_{Z'_{LP}}$width)\tabularnewline
\hline 
$1500\pm4\Gamma_{Z'}$  & $6.60\times10^{-3}$  & $1.34\times10^{-2}$  & $5.63\times10^{-3}$  & $2.33\times10^{-2}$\tabularnewline
\hline 
$2000\pm4\Gamma_{Z'}$  & $1.87\times10^{-3}$  & $3.54\times10^{-3}$  & $1.50\times10^{-3}$  & $6.72\times10^{-3}$\tabularnewline
\hline 
$2500\pm4\Gamma_{Z'}$  & $5.26\times10^{-4}$  & $1.02\times10^{-3}$  & $4.34\times10^{-4}$  & $1.97\times10^{-3}$\tabularnewline
\hline 
$3000\pm4\Gamma_{Z'}$  & $1.67\times10^{-4}$  & $3.09\times10^{-4}$  & $1.33\times10^{-4}$  & $6.23\times10^{-4}$\tabularnewline
\hline 
\end{tabular}
\end{table}

\begin{figure}
\includegraphics[scale=0.8]{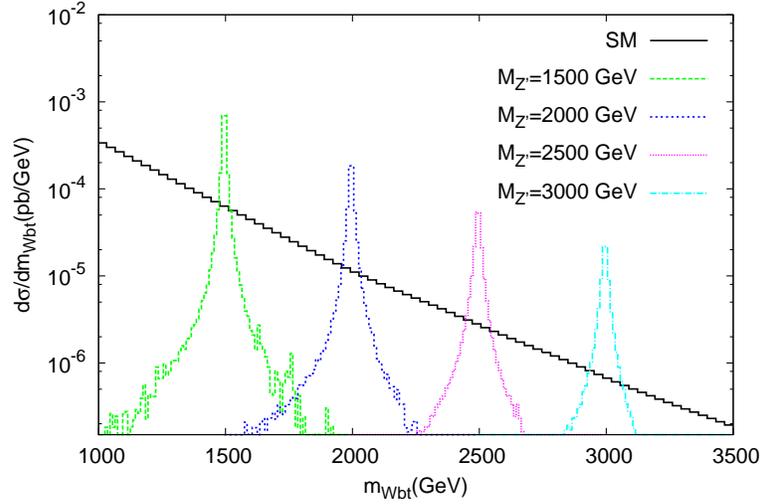}

\caption{The invariant mass distribution of the ($W^{+}b\bar{\overline{t}+W^{-}\overline{b}t}$)
system for the SM and $Z'_{\eta}$ model with different mass values
of $Z'$ boson at the LHC with $\sqrt{s}=13$ TeV. The results are
obtained for $x=0.1$ and $m_{t'}=700$ GeV. \label{fig:14} }
\end{figure}

\begin{figure}
\includegraphics[scale=0.8]{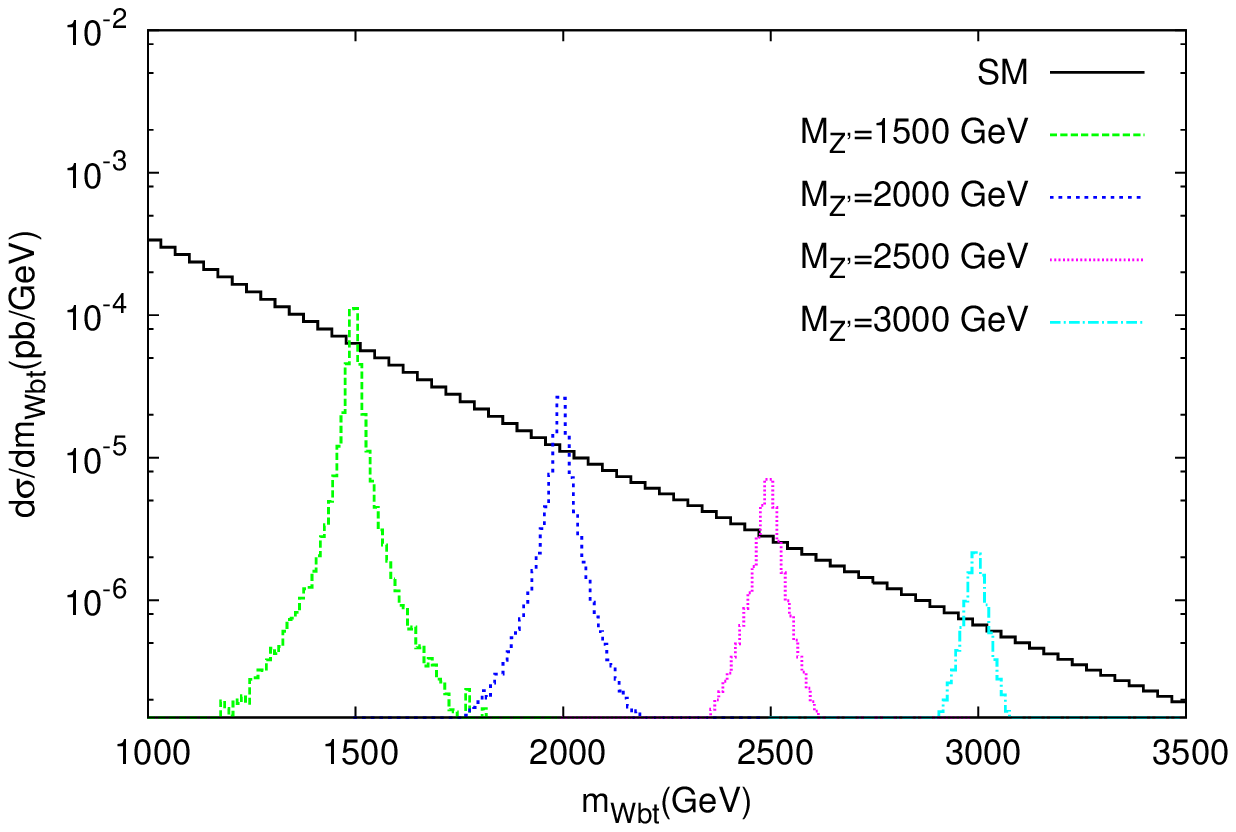}

\caption{The same as Figure 13, but for $Z'_{\chi}$ model. \label{fig:15} }
\end{figure}

\begin{figure}
\includegraphics{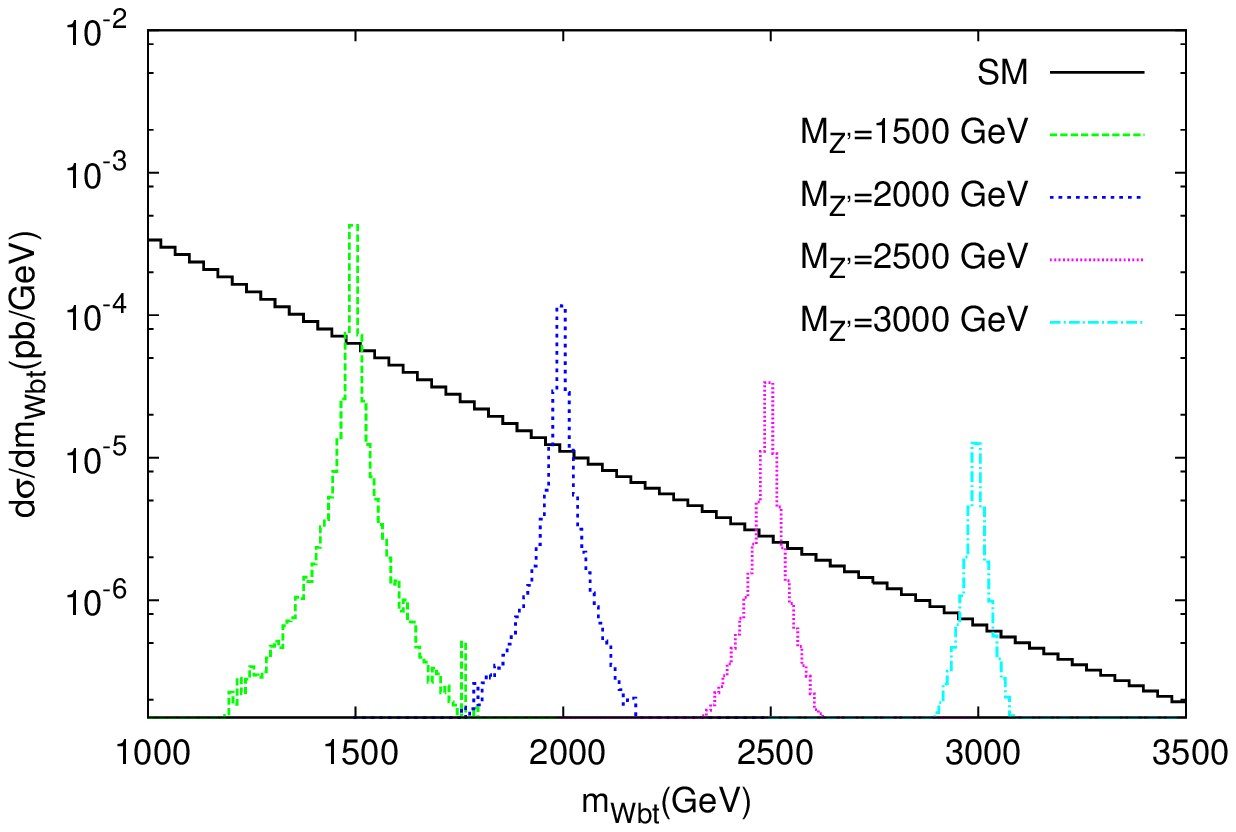}

\caption{The same as Figure 13, but for $Z'_{\psi}$ model. \label{fig:16}}
\end{figure}

\begin{figure}
\includegraphics{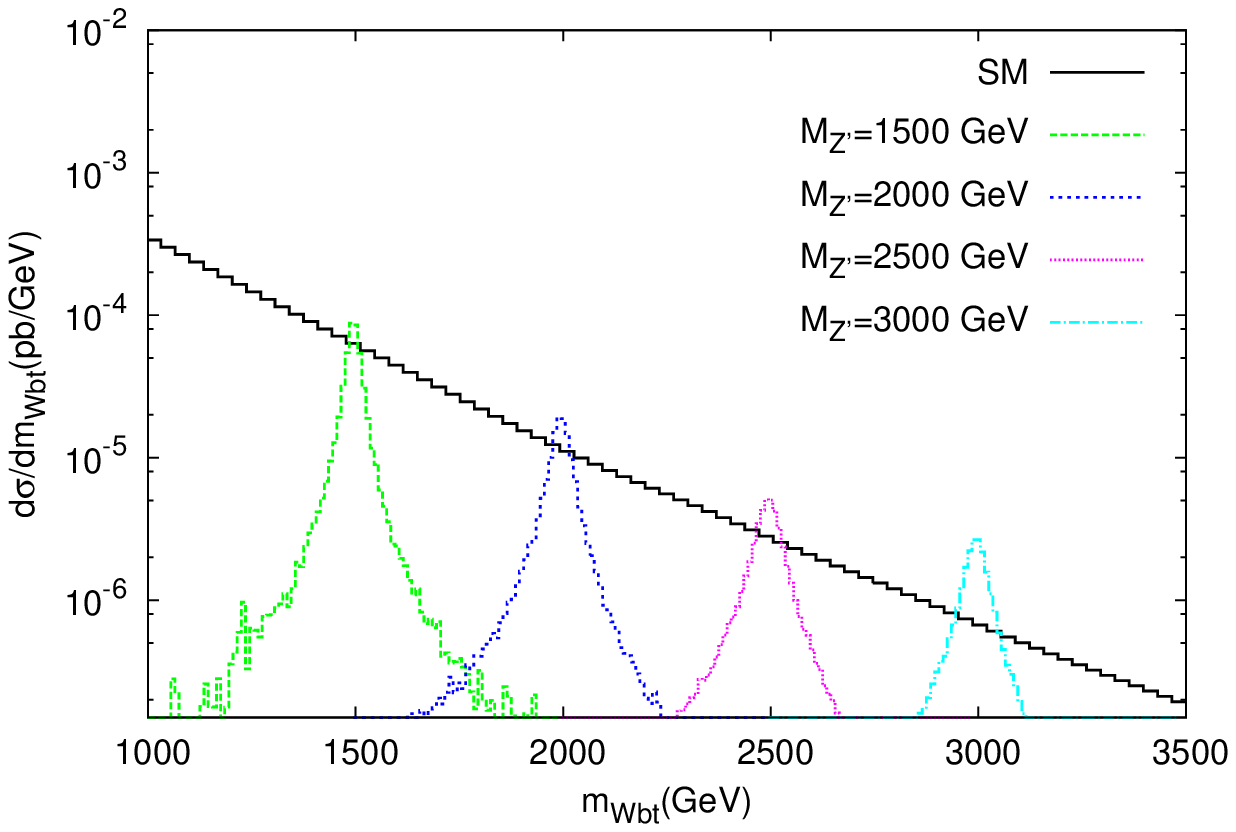}

\caption{The same as Figure 13, but for $Z'_{LP}$ model. \label{fig:17}}
\end{figure}

We plot the invariant mass distribution of the $Wbt$ system for the
signal (with $x=0.1$ and $m_{t'}=700$ GeV) and background at the
LHC with $\sqrt{s}=13$ TeV in Fig. \ref{fig:14}, Fig. \ref{fig:15},
Fig. \ref{fig:16} and Fig. \ref{fig:17} for different $Z'$ models.

\section{Analysis}

\textit{\emph{Here, we consider two types of backgrounds for the analysis.
The first one has the same final state ($Wbt$) as expected for the
signal processes and the other one (pair production of top quarks
both associated with $b$-jets) is the irreducible background and
contributes to the similar final state. }}The ratio of the cross sections
for pair production of top quarks at the $\sqrt{s}=13$ TeV and $\sqrt{s}=8$
TeV is about $3.5$. The ratio of the cross sections for process $pp\to(t'\bar{t}+t\bar{t}')+X$
is found to be about $8$ for considered $Z'$ models. It is expected
that an improvement in the statistical significance (for the center
of mass energy $\sqrt{s}=13$ TeV when compared to the case of $\sqrt{s}=8$
TeV) will be obtained. For the analysis, one can apply a high transverse
momentum ($p_{T}$) cut for the $b$-jets and the other jets. Employing
the variable $p_{T}$ cuts such that $p_{T}>100$ GeV for different
$Z'$ mass values and the rapidity cuts $|\eta|<2$ for the central
detector coverage, the results are given \textit{\emph{in Table \ref{tab:tab12},
Table \ref{tab:tab13}, Table \ref{tab:tab14} and Table \ref{tab:tab15},
we give the number of signal ($S$) and background ($B$) events by
assuming integrated luminosity of $L_{int}=10^{5}$ pb$^{-1}$ per
year. For the FCNC coupling parameter $x=0.1$, the LHC is able to
measure the $Z'$ mass up to about $3000$ GeV with the associated
productions of the new heavy quark $t'$ and top quark.}} The statistical
significance ($SS$) values for the final state are given in Table
\ref{tab:tab12} - Table \ref{tab:tab15} for different $Z'$ boson
masses.

\begin{table}
\caption{The number of signal and background events for the final state $l^{\pm}+2b_{jet}+2jet+MET$
at the center of mass energy $\sqrt{s}=13$ TeV and integrated luminosity
$L_{int}=10^{5}\mbox{ pb}^{-1}$. The numbers in the parentheses denote
the signal significances ($SS$). These results are achieved for the
$Z'_{\eta}$ model and parameter $x=0.1$. \label{tab:tab12}}

\begin{tabular}{|c|c|c|c|c|}
\hline 
\multicolumn{1}{|c|}{} & \multicolumn{3}{c|}{Signal - $\left(t'\bar{t}+\bar{t'}t\right)\to2\times(l^{\pm}+2b_{jet}+2jet+MET)$} & Background - $(W^{+}b\bar{t}+W^{-}\bar{b}t)$\tabularnewline
\hline 
$M_{Z'_{\eta}}(\mbox{GeV})$  & $m_{t'}=600\mbox{ GeV}$  & $m_{t'}=700\mbox{ GeV}$  & $m_{t'}=800\mbox{ GeV}$  & $2\times(l^{\pm}+2b_{jet}+2jet+MET)$\tabularnewline
\hline 
1500  & 305.0 (30.7)  & 277.8 (28.0)  & 245.6 (24.7)  & 98.6\tabularnewline
\hline 
2000  & 91.2 (17.3)  & 68.8 (16.4)  & 81.0 (15.3)  & 28.0\tabularnewline
\hline 
2500  & 30.0 (10.6)  & 29.2 (10.3)  & 28.0 (9.9)  & 7.8\tabularnewline
\hline 
3000  & 12.4 (7.9)  & 12.4 (7.8)  & 12.0 (7.6)  & 2.6\tabularnewline
\hline 
\end{tabular}
\end{table}

\begin{table}
\caption{The same as Table \ref{tab:tab12}, but for $Z'_{\chi}$. \label{tab:tab13}}

\begin{tabular}{|c|c|c|c|c|}
\hline 
\multicolumn{1}{|c|}{} & \multicolumn{3}{c|}{Signal - $\left(t'\bar{t}+\bar{t'}t\right)\to2\times(l^{\pm}+2b_{jet}+2jet+MET)$} & Background - $\left(W^{+}b\bar{t}+W^{-}\bar{b}t\right)$\tabularnewline
\hline 
$M_{Z'_{\chi}}(\mbox{GeV})$  & $m_{t'}=600\mbox{ GeV}$  & $m_{t'}=700\mbox{ GeV}$  & $m_{t'}=800\mbox{ GeV}$  & $2\times(l^{\pm}+2b_{jet}+2jet+MET)$\tabularnewline
\hline 
1500  & 68.0 (4.8)  & 61.6 (4.4)  & 54.8 (4.0)  & 200.4\tabularnewline
\hline 
2000  & 19.8 (2.7)  & 18.8 (2.5)  & 17.4 (2.4)  & 53.0\tabularnewline
\hline 
2500  & 6.2 (1.6)  & 6.0 (1.6)  & 5.8 (1.6)  & 15.2\tabularnewline
\hline 
3000  & 2.2 (1.4)  & 2.2 (1.4)  & 2.0 (1.4)  & 4.6\tabularnewline
\hline 
\end{tabular}
\end{table}

\begin{table}
\caption{The same as Table \ref{tab:tab12}, but for $Z'_{\psi}$. \label{tab:tab14}}

\begin{tabular}{|c|c|c|c|c|}
\hline 
\multicolumn{1}{|c|}{} & \multicolumn{3}{c|}{Signal - $\left(t'\bar{t}+\bar{t'}t\right)$$\rightarrow2\times\left(l^{\pm}+2b_{jet}+2jet+MET\right)$} & Background - $\left(W^{+}b\bar{t}+W^{-}\bar{b}t\right)$\tabularnewline
\hline 
$m_{Z'_{\psi}}(\mbox{GeV})$  & $m_{t'}=600\mbox{ GeV}$  & $m_{t'}=700\mbox{ GeV}$  & $m_{t'}=800\mbox{ GeV}$  & $2\times(l^{\pm}+2b_{jet}+2jet+MET)$\tabularnewline
\hline 
1500  & 176.5 (19.2)  & 158.6 (17.3)  & 133.6 (14.6)  & 84.2\tabularnewline
\hline 
2000  & 51.3 (10.8)  & 49.4 (10.4)  & 46.4 (9.8)  & 22.4\tabularnewline
\hline 
2500  & 16.6 (6.5)  & 16.3 (6.4)  & 15.7 (6.2)  & 6.5\tabularnewline
\hline 
3000  & 6.5 (4.6)  & 6.4 (4.5)  & 6.3 (4.6)  & 2.0\tabularnewline
\hline 
\end{tabular}
\end{table}

\begin{table}
\caption{The same as Table \ref{tab:tab12}, but for $Z'_{LP}$. \label{tab:tab15}}

\begin{tabular}{|c|c|c|c|c|}
\hline 
\multicolumn{1}{|c|}{} & \multicolumn{3}{c|}{Signal - $\left(t'\bar{t}+\bar{t'}t\right)$$\rightarrow2\times\left(l^{\pm}+2b_{jet}+2jet+MET\right)$} & Background - $\left(W^{+}b\bar{t}+W^{-}\bar{b}t\right)$\tabularnewline
\hline 
$M_{Z'_{LP}}(\mbox{GeV})$  & $m_{t'}=600\mbox{ GeV}$  & $m_{t'}=700\mbox{ GeV}$  & $m_{t'}=800\mbox{ GeV}$  & $2\times(l^{\pm}+2b_{jet}+2jet+MET)$\tabularnewline
\hline 
1500  & 77.3 (4.1)  & 71.5 (3.8)  & 64.3 (3.4)  & 348.5\tabularnewline
\hline 
2000  & 22.9 (2.3)  & 22.0 (2.2)  & 20.8 (2.1)  & 100.5\tabularnewline
\hline 
2500  & 7.6 (1.4)  & 7.4 (1.4)  & 7.2 (1.3)  & 29.5\tabularnewline
\hline 
3000  & 3.7 (1.2)  & 3.7 (1.2)  & 3.7 (1.2)  & 9.3\tabularnewline
\hline 
\end{tabular}
\end{table}

\textit{\emph{In the analysis, we reconstruct the invariant mass of
$Wbt$ system around the $Z'$ boson mass which are shown in Fig.}}
\ref{fig:14}, \textit{\emph{Fig.}} \ref{fig:15}, \textit{\emph{Fig.}}
\ref{fig:16},\textit{\emph{ Fig.}} \ref{fig:17}. \textit{\emph{We
assume top quark decay $t(\bar{t})\to W^{+}b(W^{-}\bar{b})$, where
the $W$ boson can decay}} leptonically or hadronically. \textit{\emph{In
the final state $W^{+}W^{-}b\bar{b}$, we assume the $b$-tagging
efficiency as $50\%$ for each of the $b$-quarks.}} We take into
account the channel in which one of the $W$ bosons decay leptonically,
while the other decays hadronically. We calculate the cross section
of the background in the mass bin widths for each $M_{Z'}$ value;
as an example of the $Z'_{\eta}$ model, for the $M_{Z'}=1500$ GeV
we take the invariant mass interval $\Delta m_{Wbt}\simeq40$ GeV,
and we find the background cross section $\sigma_{B}=6.60\times10^{-3}$
pb for process $pp\to(W^{-}\bar{b}t+W^{+}b\bar{t})+X$.

\begin{figure}
\includegraphics[scale=0.8]{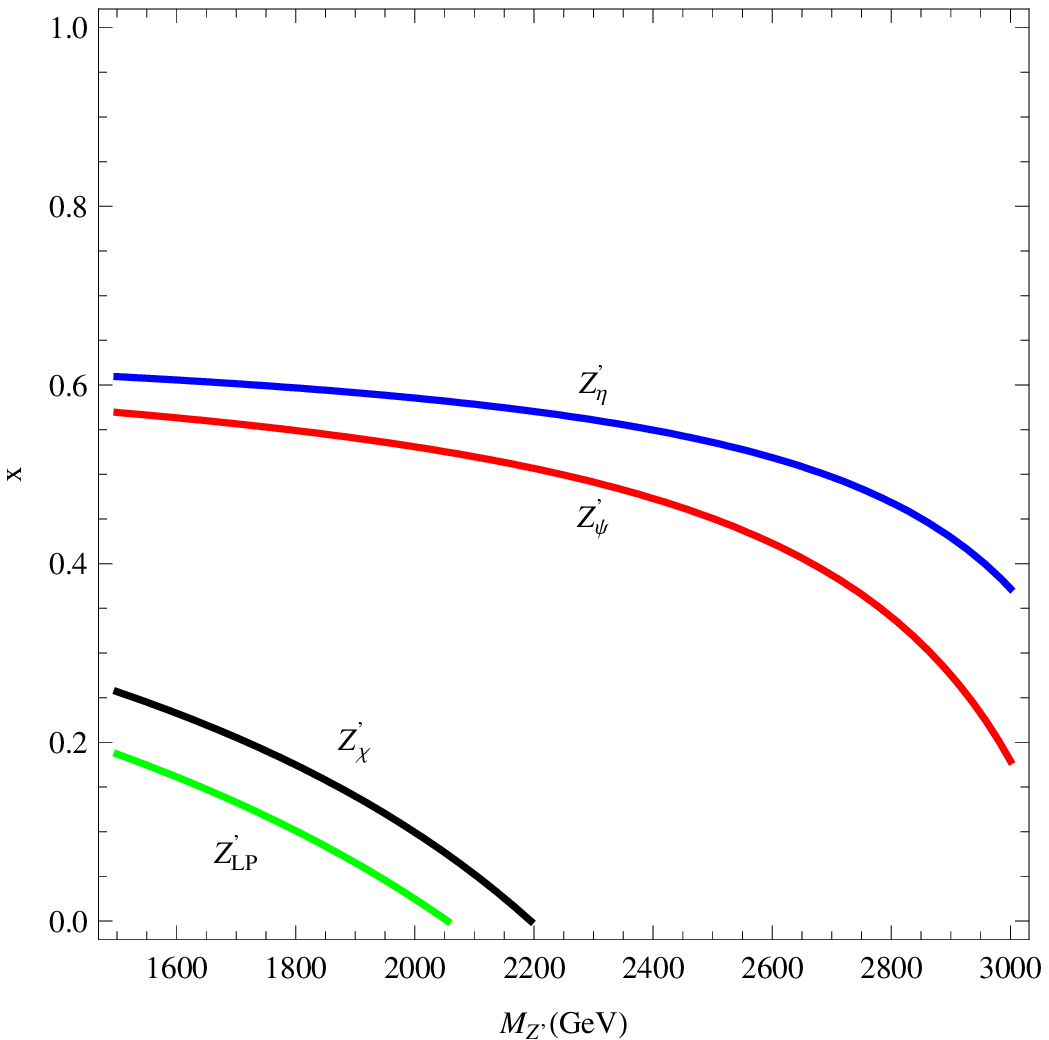}

\caption{The contour plot for the evidence of $Z'$ boson at the LHC ($\sqrt{s}=13$
TeV) with $L_{int}=100$ fb$^{-1}$. The parameters used in the analysis
are explained in the text. \label{fig:fig18}}
\end{figure}

We plot the observability contours in the plane of model parameters
$(x-M_{Z'})$, for different $Z'$ models as shown in Fig. \ref{fig:fig18}
at the LHC with $\sqrt{s}=13$ TeV and $L_{int}=100$ fb$^{-1}$.
The curves with labels $Z'_{LP}$, $Z'_{\chi}$, $Z'_{\psi}$ and
$Z'_{\eta}$ show the accessible regions (below the curves) of the
model parameters at the LHC. For the $Z'_{\eta}$ model, the FCNC
parameter bounds from $x=0.6-0.4$ can be searched for the mass range
of $M_{Z'}=1500-3000$ GeV.

\section{Conclus\i on}

We consider the associated productions of new heavy quark $t'$ and
top quark (with the subsequent decay channel $t'\rightarrow W^{+}b$)
through the $Z'$ exchange diagrams at the LHC. We find the discovery
regions of the parameter space for the single productions of new heavy
quarks through FCNC interactions with the new $Z'$ boson. In the
models considered in this paper, the single production of new heavy
quarks at the LHC can have the contributions from the couplings of
$Z'q\bar{q}$ and the FCNC couplings of $Z'q\bar{q}'$ (where $q,q'=u,c,t,t'$).
For the FCNC parameter range ($0<x<1$ means maximal to minimal FCNC)
the LHC can have the potential to produce new heavy quarks which couple
to the $Z'$ boson predicted by specific models. 
\begin{acknowledgments}
O.C's work is supported in part by the Turkish Atomic Energy Authority
(TAEK) under the project Grant No. 2011TAEKCERN-A5.H2.P1.01-19. 
\end{acknowledgments}

\end{document}